\documentclass[aps,prd,twocolumn,notitlepage,superscriptaddress,nofootinbib]{revtex4-1}
\usepackage[utf8]{inputenc}
\usepackage[english]{babel}
\usepackage{amsmath}
\usepackage{amsfonts}
\usepackage{amssymb}
\usepackage{listings}
\usepackage{lipsum}
\usepackage{multirow}
\usepackage{datetime}
\usepackage{graphicx}
\usepackage{mathtools}
\usepackage{mathrsfs}
\usepackage{dcolumn}
\usepackage{multirow}
\usepackage{color}   %May be necessary if you want to color links
\usepackage{hyperref}
\hypersetup{
    colorlinks=true, 
    pdfborder = {0 0 0.5 [3 3]},
    anchorcolor=black,
    citecolor=blue,
    linktoc=all,    
    linktocpage=true,
    linkcolor=red,
	urlcolor=blue
}

\begin{document}

\title{\texttt{SuperRad}: Modeling the black hole superradiance gravitational waveform}
\author{Nils Siemonsen}
\email[]{nsiemonsen@perimeterinstitute.ca}
%\homepage[]{Your web page}
%\thanks{}
%\altaffiliation{}
\affiliation{Perimeter Institute for Theoretical Physics, Waterloo, Ontario N2L 2Y5, Canada}
\affiliation{Arthur B. McDonald Canadian Astroparticle Physics Research Institute, 64 Bader Lane, Queen's University, Kingston, Ontario K7L 3N6, Canada}
\affiliation{Department of Physics \& Astronomy, University of Waterloo, Waterloo, Ontario N2L 3G1, Canada}
\author{Taillte May}
\affiliation{Perimeter Institute for Theoretical Physics, Waterloo, Ontario N2L 2Y5, Canada}
\affiliation{Department of Physics \& Astronomy, University of Waterloo, Waterloo, Ontario N2L 3G1, Canada}
\author{William E.\ East}
\email[]{weast@perimeterinstitute.ca}
\affiliation{Perimeter Institute for Theoretical Physics, Waterloo, Ontario N2L 2Y5, Canada}

\date{\today}

\begin{abstract} 
Gravitational signatures of black hole superradiance are a unique probe of
ultralight particles that are weakly-coupled to ordinary matter.  The existence
of an ultralight boson would lead spinning black holes with size comparable to
the Compton wavelength of the boson to become superradiantly unstable to forming
an oscillating cloud, spinning down the black hole, and radiating gravitational
waves in the process.  However, maximizing the chance of observing
such signals or, in their absence, placing the strongest constraints on the
existence of such particles, requires accurate theoretical predictions.  In
this work, we introduce a new gravitational waveform model, \texttt{SuperRad},
that models the dynamics, oscillation frequency, and gravitational wave
signals of these clouds by 
combining numerical results in the relativistic regime with fits
calibrated to analytical estimates, covering
the entire parameter space of ultralight scalar and vector clouds with the lowest
two azimuthal numbers ($m=1$ and 2). 
We present new calculations of the gravitational wave frequency evolution
as the boson cloud dissipates, including using fully general-relativistic
methods to quantify the error in more approximate treatments.
Finally, as a first application, we
assess the viability of conducting follow-up gravitational wave searches for
ultralight vector clouds around massive black hole binary merger remnants. We
show that LISA may be able to probe vector masses in the range from $1\times
10^{-16}$ eV to $6\times 10^{-16}$ eV using follow-up gravitational wave
searches.  
\end{abstract}

\maketitle

\section{Introduction}

The advent of gravitational wave (GW) astronomy has brought a powerful tool to probe
new physics in regimes that have been inaccessible to previous experiments. Illusive,
weakly-coupled ultralight bosons beyond the Standard Model of particle physics
have been conjectured to solve various problems in high energy physics and
cosmology. However, terrestrial experiments require sufficiently strong coupling
to the Standard Model for a direct detection. Therefore, gravitational
signatures, which assume only that these ultralight particles gravitate, are
ideal for efficiently probing the weak-coupling parameter space inaccessible
to other observational efforts. Namely, black hole (BH) superradiance provides
a purely gravitational mechanism through which ultralight bosonic particles extract rotational
energy from spinning BHs with observable consequences. 

Bosonic waves whose frequency satisfy the superradiance condition are amplified
when scattering off a rotating BH
\cite{Starobinsky:1973aij,1971JETPL..14..180Z}, extracting rotational energy in
a type of Penrose process \cite{Penrose:1971uk}. If the underlying bosonic
particle is massive, there is an instability associated with superradiance, leading 
to the formation of exponentially growing, oscillating bound states---
superradiant clouds--- around the BH. Assuming self-interactions and couplings to other
matter are sufficiently weak, the instability saturates gravitationally as the
BH loses energy and angular momentum, and is spun down.  At this point, the
system transitions from an exponentially growing phase to a phase characterized
by quasi-monochromatic GW emission which causes the cloud to
slowly dissipate.  Therefore, the presence of the superradiant cloud leaves
observational signatures in the BH spin distribution and the GW emission. 

This observational window allows us to probe various well-motivated extensions
to the Standard Model \cite{Arvanitaki:2009fg}. For scalar bosons, the QCD
axion (solving the strong CP-problem), axion dark matter (solving the dark
matter problem), and various quantum gravity motivated axion-like particles,
are ultralight candidates capable of forming superradiant clouds
\cite{Peccei:1977hh,Weinberg:1977ma,Wilczek:1977pj,Arvanitaki:2009fg,Essig:2013lka,Hui:2016ltb,Marsh:2015xka}.
The dark photon is a viable candidate to make up a significant fraction of the
dark matter, or could emerge in the low-energy limits of quantum gravity
\cite{Goodsell:2009xc,Jaeckel:2010ni}. Ultralight spin-2 fields are a possible
modification of general relativity \cite{Clifton:2011jh}. Hence, a wide variety of
models could be constrained, or discovered via this observational window; addressing fundamental questions in particle physics, cosmology, and high
energy physics.

To leverage the observational potential of ground- and space-based GW
detectors, accurate predictions for the involved spin-down timescales, as well
as GW frequency and amplitudes are required. Much effort has gone into
determining these for scalar bosons
\cite{Ternov:1978gq,Zouros:1979iw,Detweiler:1980uk,Dolan:2007mj,Arvanitaki_precision,Yoshino:2013ofa,Arvanitaki1,Arvanitaki2,Brito:2014wla,Yoshino:2015nsa},
vector bosons
\cite{Rosa:2011my,Pani:2012vp,Pani:2012bp,Cardoso:2018tly,Baryakhtar:2017ngi,East:2017mrj,East:2018glu,Baumann:2019eav,Siemonsen:2019ebd},
and spin-2 fields \cite{Brito:2013wya,Brito:2020lup} (see
Ref.~\cite{brito_review} for a review). Scalar bosons exhibit the longest
spin-down timescales, as well as weakest and longest GW signal after cloud
formation. Vector bosons, on the other hand, are amplified more efficiently,
leading to faster cloud growth rates and stronger, but hence shorter, GW
emissions. In modified theories of gravity, massive spin-2 fields grow the
fastest around BHs. 

Using these results, various search strategies have been employed to constrain
parts of the ultralight boson parameter space. Electromagnetic spin
measurements of stellar mass and supermassive BHs
\cite{McClintock:2013vwa,Miller:2014aaa,Reynolds:2013qqa} have been used to
disfavor ultralight scalars
\cite{Arvanitaki1,Cardoso:2018tly,brito,Stott:2020gjj} and vectors
\cite{Baryakhtar:2017ngi} in certain mass ranges.  
Similarly, measurements of
spins of the constituents of inspiraling binary BHs
\cite{Venumadhav:2019lyq,LIGOScientific:2018mvr,LIGOScientific:2018jsj} and BH population properties were
used in Ref.~\cite{Ng:2020ruv,Ng:2019jsx,Payne:2021ahy} to exclude a small scalar mass range. Stochastic
GW searches from a population of BH-cloud systems were used to
constrain the scalar \cite{brito,Tsukada:2019,brito_short} and vector
\cite{Tsukada:2020lgt} masses, while various directed and blind continuous GW
search techniques lead to constraints
\cite{Sun:2019mqb,allsky_sr_search,allsky_sr_method,Isi:2018pzk,CW_galactic,KAGRA:2022osp,KAGRA:2021tse,Dergachev:2019wqa}. The
presence of cloud around a constituent BH within a binary could also affect the
inspiral dynamics, leaving observable signatures in the emitted GW
waveform~\cite{Hannuksela:2018izj,Baumann:2019eav,Baumann:2021fkf,Choudhary:2020pxy}.

The methods used to determine the observable consequences of superradiance for a
given BH of mass $M$, are classified by the
their regime of validity for the dimensionless gravitational fine structure constant
\begin{align}
\alpha\approx 0.075\left(\frac{M}{10\ M_\odot}\right)\left(\frac{\mathcal{M}}{10^{-12}\ \text{eV}}\right),
\label{eq:alphadefinition}
\end{align}
where $\mathcal{M}$ is the mass of the ultralight particle.  Analytic
techniques are most accurate for $\alpha\ll 1$, the regime where the boson
cloud is farther away from the BH and can be treated non-relativistically.
However, numerical approaches are required for systems with
$\alpha\sim\mathcal{O}(1)$, where the cloud sits close to the BH, and relativistic
effects are import.  Analytic estimates have been pushed to high orders in an
expansion around small $\alpha$, while numerical techniques have been refined
to include large parts of the parameter space.  Despite this progress, and the
significant impact of gravitational probes, most gravitational and
electromagnetic wave search campaigns for signatures of BH superradiance have
employed lower-order, potentially inaccurate, predictions, leaving the most
favorable parts of the parameter space unexplored. 

Here, we introduce \texttt{SuperRad}, an open source BH superradiance waveform
model incorporating state-of-the-art theoretical predictions for BH spin-down
and GW observables across the entire relevant parameter space
in a simple, ready-to-use \texttt{python} package\footnote{Available at \url{www.bitbucket.org/weast/superrad} .}.  A primary
goal is to provide a tool to efficiently and accurately interpret GW search
results of current and future ground- and space-based GW 
observatories.  As part of this work, we present new calculations of the
frequency evolution of the boson cloud oscillations and attendant GWs 
due to the changing mass of the boson cloud. We compare the analytic
frequency evolution in the non-relativistic limit to both approximate
quasi-relativistic calculations, as well as fully general-relativistic ones, to
determine their accuracy in the relativistic regime.

As a first application, we use \texttt{SuperRad} to show that the Laser
Interferometer Space Antenna (LISA) should in principle be able to probe
ultralight boson masses from $1\times10^{-16}$ eV to $6\times 10^{-16}$ eV by
performing follow-up searches for GWs from boson clouds arising
around the remnants of massive BH binary mergers. 
Such follow-up searches have
been previously discussed in the context of stellar mass BH
mergers~\cite{Ghosh:2018gaw,Isi:2018pzk,Chan:2022dkt}, and are especially promising because
the observation of the binary BH merger waveform gives definitive
information on the properties of the remnant BH, allowing one to place
constraints in the absence of a signal without further assumptions. By contrast, 
other search methods outlined above rely on further assumptions and are subject to various uncertainties: electromagnetic spin measurements are contingent on astrophysical uncertainties and may be invalidated by weak couplings of the ultralight boson to the Standard Model \cite{Siemonsen:2022ivj}, spin measurements of constitutents of inspiraling binary BHs have large statistical uncertainties, and constraints based on blind continuous waves and stochastic gravitational wave background searches rely on BH population assumptions.

We begin in Sec.~\ref{sec:example} by providing a broad overview over the expected GW signals from BH superradiance of scalar and vector clouds. In Sec.~\ref{sec:cloudproperties}, we discuss in detail how \texttt{SuperRad} determines the cloud's oscillation frequency and the superradiance instability timescales. Furthermore, we analyze the frequency shift due to the finite self-gravity of the cloud around the BH in Sec.~\ref{sec:freqshift} using Newtonian, quasi-relativistic, and fully relativistic approaches. The GW amplitude and waveform is discussed in Sec.~\ref{sec:gw}. Following this, we outline the linear evolution of the cloud as well as the accompanying GW signature in Sec.~\ref{sec:cloudevolution}, and close with the application of \texttt{SuperRad} to analyze the prospects of follow-up searches with LISA in Sec.~\ref{sec:lisafollowup}. We use $G=c=1$ units throughout.

\section{Overview and Example} \label{sec:example}
\begin{figure}
\includegraphics[width=0.49\textwidth]{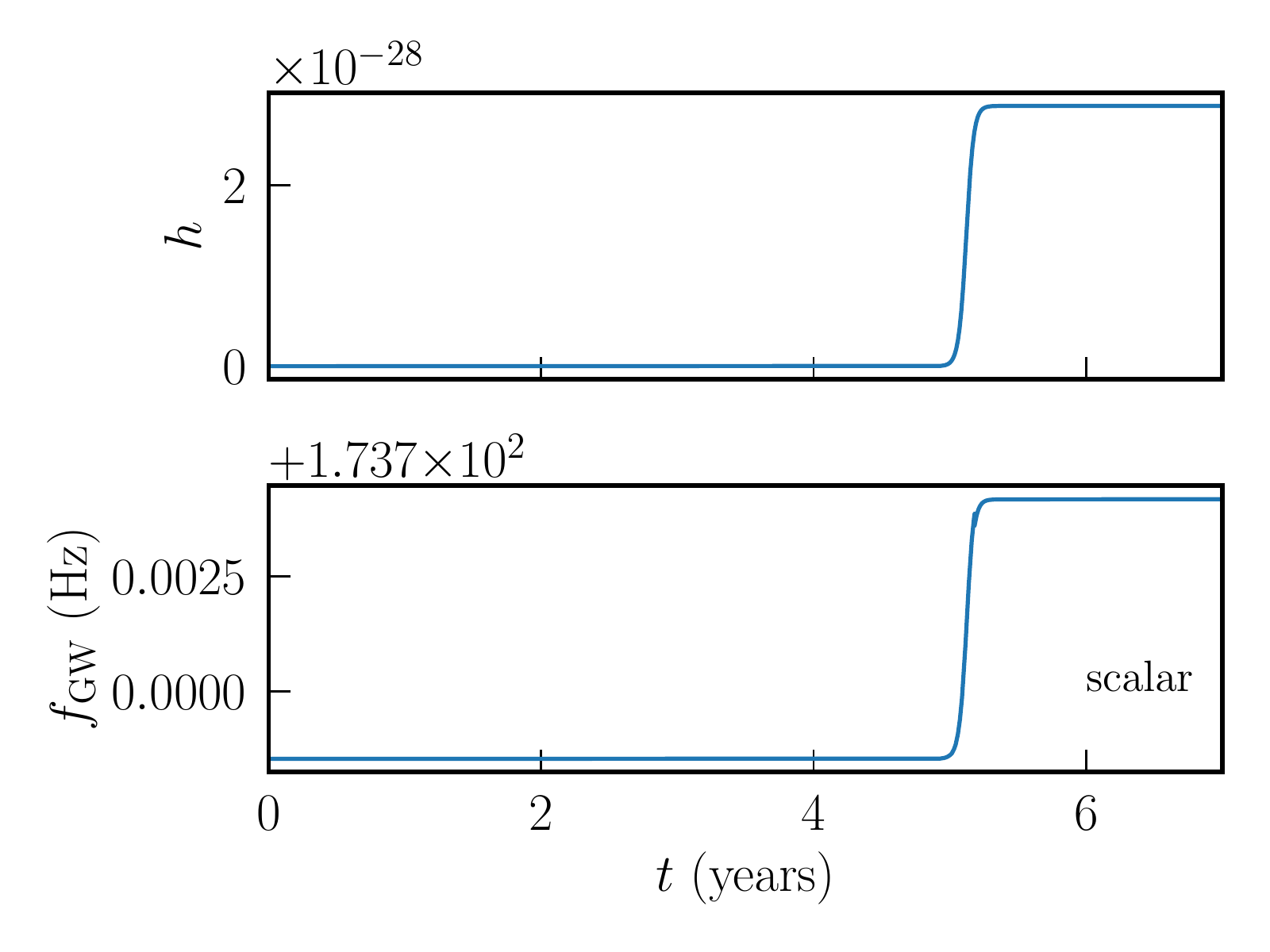}
\includegraphics[width=0.49\textwidth]{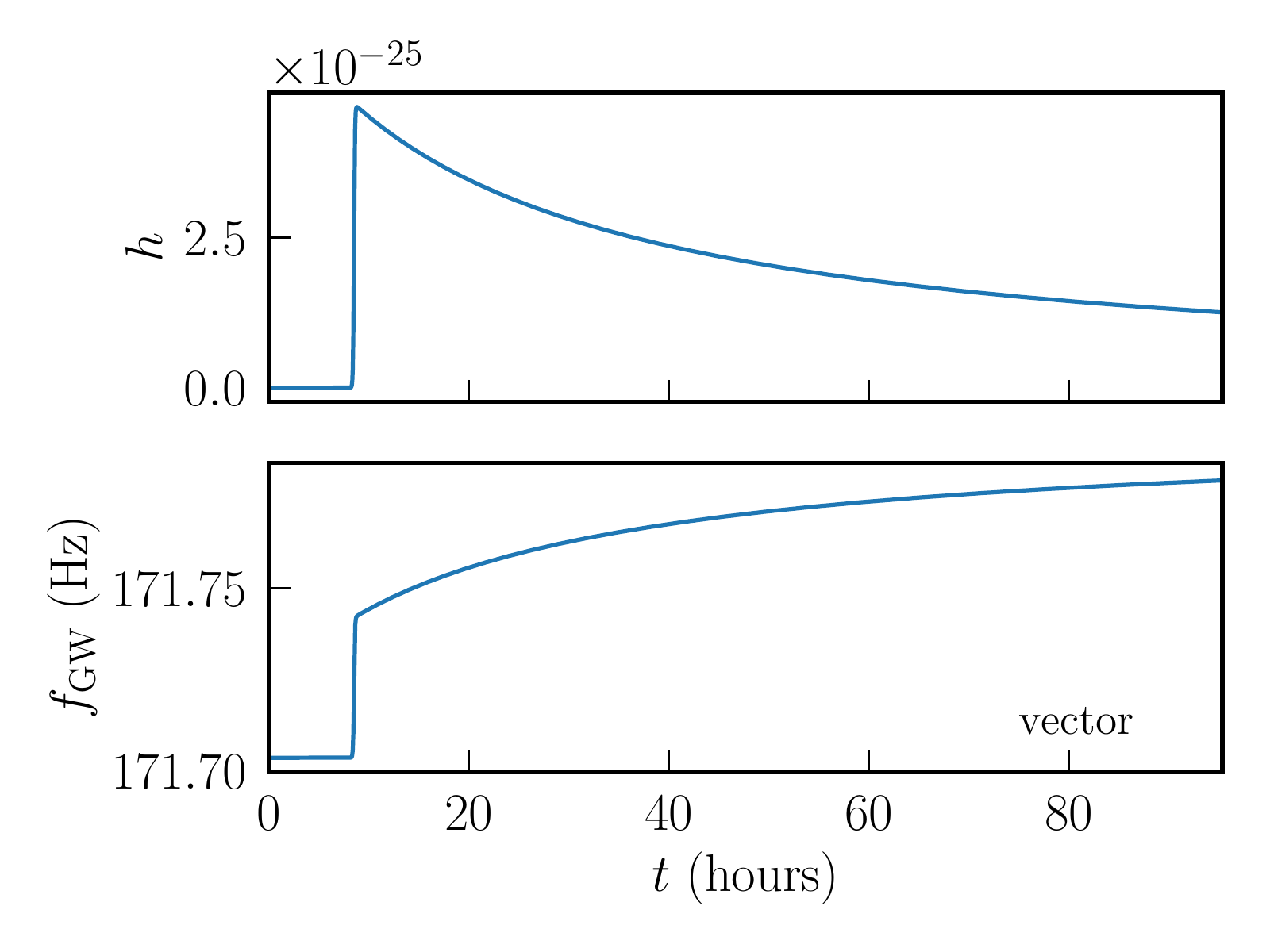}
\caption{
    The GW strain $h$ and frequency $f_{\rm GW}$ as a function of time for a
    BH with $M=62\ M_{\odot}$ and $a_*=0.67$ at a distance of 
    410 Mpc subject to the superradiant instability of a boson with
    mass $3.6\times10^{-13}$ eV.
    The top set of panels shows the scalar boson case,
    while the bottom set shows the vector case. Note the difference
    in timescales shown, since in the scalar (vector) case the cloud grows 
    on timescales of $\sim 5$ years (9 hours) and decays through GW radiation on timescales of 
    $\sim 9000$ years (1 day).
    Time is measured since the BH was formed, assuming the cloud started as a single boson.
    }    
\label{fig:strain_freq}
\end{figure}

We begin with an example to illustrate the expected GW signal
from superradiant clouds, and give an overview of the different effects that go into calculating it.
We consider parameters consistent with
the remnant from a GW150914-like binary BH merger. In particular, we
consider a BH with mass $M=62\ M_{\odot}$ and dimensionless spin\footnote{Defined by the ratio of angular momentum $J$ to the mass square of the BH: $a_*=J/M^2$.} $a_*=0.67$ at a distance of 410
Mpc~\cite{LIGOScientific:2016vlm} and determine the resulting GW signal if
there were an ultralight boson---scalar or vector---with mass
$\mathcal{M}=3.6\times10^{-13}$ eV (hence $\alpha\approx0.17$). For
simplicity, here we assume the angular momentum points in the direction of the
observer---hence both GW polarizations are equal---and ignore redshift
effects. The GW strain and frequency calculated with \texttt{SuperRad} for
both the scalar boson case and the vector case are shown in
Fig.~\ref{fig:strain_freq}.

There are a number of different parts that go into these calculations. First,
one determines the superradiant instability timescale by solving for the
fastest growing mode of the massive scalar or vector equations of motion on the
BH spacetime as described in Sec.~\ref{sec:cloudproperties}.  This gives the
timescale over which the boson cloud mass, and hence the GW 
amplitude, grows exponentially in time.  From Fig.~\ref{fig:strain_freq}, it
can be seen that the $e$-folding time of the cloud mass (half the
$e$-folding time of the field $\tau_I$) is much slower for the scalar case
($\tau_I/2\sim 10$ days) compared to the vector case ($\tau_I/2\sim 3$
minutes). Taking into account the resulting decrease in the mass and spin of
the BH as the boson cloud grows, as described in Sec.~\ref{sec:cloudevolution},
the instability timescale becomes longer and longer as the horizon frequency of
the BH approaches the oscillation frequency of the cloud.  As the instability
saturates, and the cloud mass reaches its maximum value, the dissipation of the
cloud through gravitational radiation becomes dominant, leading to a slow
decrease in cloud mass. The rate at which energy is lost through gravitational
radiation $P_{\rm GW}$, as well as the two strain polarizations $h_+$ and
$h_\times$, are calculated by solving for linearized metric perturbations on
the BH spacetime, sourced by the oscillating cloud solution, as described in
Sec.~\ref{sec:gw}. As can be seen in Fig.~\ref{fig:strain_freq}, in
the scalar case the decay of GW amplitude is negligible on any reasonable
observing timescale, taking on the order of $10^4$ years, while in the vector
case, the cloud mass and GW amplitude decrease on timescales of
days.

The gravitational frequency shown in Fig.~\ref{fig:strain_freq} exhibits
an increase or ``chirp" in frequency, first as the BH loses mass and  
the cloud grows exponentially, and then more slowly as the boson cloud dissipates. 
Calculating this frequency shift
requires accounting for the self-gravity of the boson cloud, which slightly
red-shifts the oscillation frequency of the cloud, and hence the gravitational
waves (which have twice the frequency of the cloud oscillations), as described in
Sec.~\ref{sec:freqshift}.  Though the change in frequency is small, because the
GW signal persists for many cycles, this is still an important
effect.

\section{Cloud properties} \label{sec:cloudproperties}

In this section, we outline the superradiant cloud properties relevant for
observational signatures such as BH spin-down or GW emission. This includes a
brief discussion of how estimates for the superradiant instability timescale
$\tau_I$ and the emitted GW frequency $f_{\rm GW}$ are obtained for different values of the BH mass,
spin, and the gravitational fine structure constant $\alpha$. We
defer the analysis of the dependency of the cloud frequency on cloud mass, 
and the cloud dynamics to
Secs.~\ref{sec:freqshift} and \ref{sec:cloudevolution}, respectively.
\texttt{SuperRad} combines analytic and numerical predictions, valid for
$\alpha\ll 1$ and $\alpha\sim\mathcal{O}(1)$, and utilizes numerically
calibrated higher-order expansions to interpolate between the two regimes.

In most of the following calculations, we assume a fixed Kerr BH spacetime
$g_{\mu\nu}$, and consider scalar and vector bosonic fields, as well as linear
metric (GW) perturbations on this background. The exception to
this is the calculation of the frequency shift due to the self-gravity of the boson-cloud.
We will discuss the validity of this
assumption further in Secs.\ref{sec:gwpowerstrain} and \ref{sec:freqshift}.
Furthermore, we neglect field self-interactions and non-minimal couplings to
the Standard Model throughout, which have been investigated in
Refs.~\cite{Yoshino:2012kn,Fukuda:2019ewf,Baryakhtar:2020gao,East:2022ppo,East:2022rsi,Omiya:2022gwu}.
Depending on the coupling strength, these can alter the superradiance dynamics.
However, here we assume that we are in the weak coupling limit, which reduces
to the purely gravitational case. Therefore, the relevant field equations to solve in
order to obtain the desired observables are the scalar and vector massive wave
equations on the spacetime
$g_{\mu\nu}$, which are given by
\begin{align}
(\square_g -\mu_S^2)\Phi=0, & & \nabla_\mu F^{\mu\nu}=\mu^2_VA^\nu,
\label{eq:fieldeq}
\end{align}
where $\mathcal{M}_S=\hbar\mu_S$ and $\mathcal{M}_V=\hbar \mu_V$ 
are the scalar and vector boson masses, respectively. Due to various symmetries of the Kerr spacetime, solutions to the field equations \eqref{eq:fieldeq} can be written in the form
\begin{align}
A_\mu,\Phi\sim e^{-i(\omega t-m\varphi)},
\label{eq:fieldansatz}
\end{align}
where we introduced the azimuthal mode number $m$ and complex frequency $\omega$. Here, and in the following,
we refer to the Boyer-Lindquist time, radius, polar and azimuthal coordinate as
$t$, $r$, $\theta$, and $\varphi$. Without loss of generality, we assume the
azimuthal index to satisfy $m\geq 0$ throughout. Lastly, we label all
quantities defined both for scalar and vector fields with $\sigma\in\{S,V\}$.
The fields $A_\mu$ and $\Phi$ are susceptible to the superradiance instability,
if the superradiance condition
\begin{align}
0<\omega_R <m\Omega_H,
\end{align}
is satisfied, where $\Omega_H$ is the horizon frequency of the BH. In the
ansatz \eqref{eq:fieldansatz}, the frequency $\omega=\omega_R+i\omega_I$
encodes both the oscillation frequency of the cloud,
which is half of the characteristic GW frequency
$f_{\rm GW}=2\omega_R/(2\pi)$ (up to self-gravity corrections), and the instability growth timescale
$\tau_I=1/\omega_I$. For fixed mode number $m_\sigma$, these observables (in units
of $M$) depend
only on $\alpha$ and spin $a_*$, i.e., $\omega(\alpha,a_*)$. 

In what follows, we begin by outlining \texttt{SuperRad's} coverage of
the $(\alpha,a_*)$ parameter
space in Sec.~\ref{sec:parameterspace}, and then 
we illustrate
how analytic and numerical results are used to
calibrate \texttt{SuperRad} in the intermediate regime
in Secs.~\ref{sec:frequency} and \ref{sec:timescale}.

\subsection{Parameter space} \label{sec:parameterspace}

\begin{figure}[t]
\includegraphics[width=0.49\textwidth]{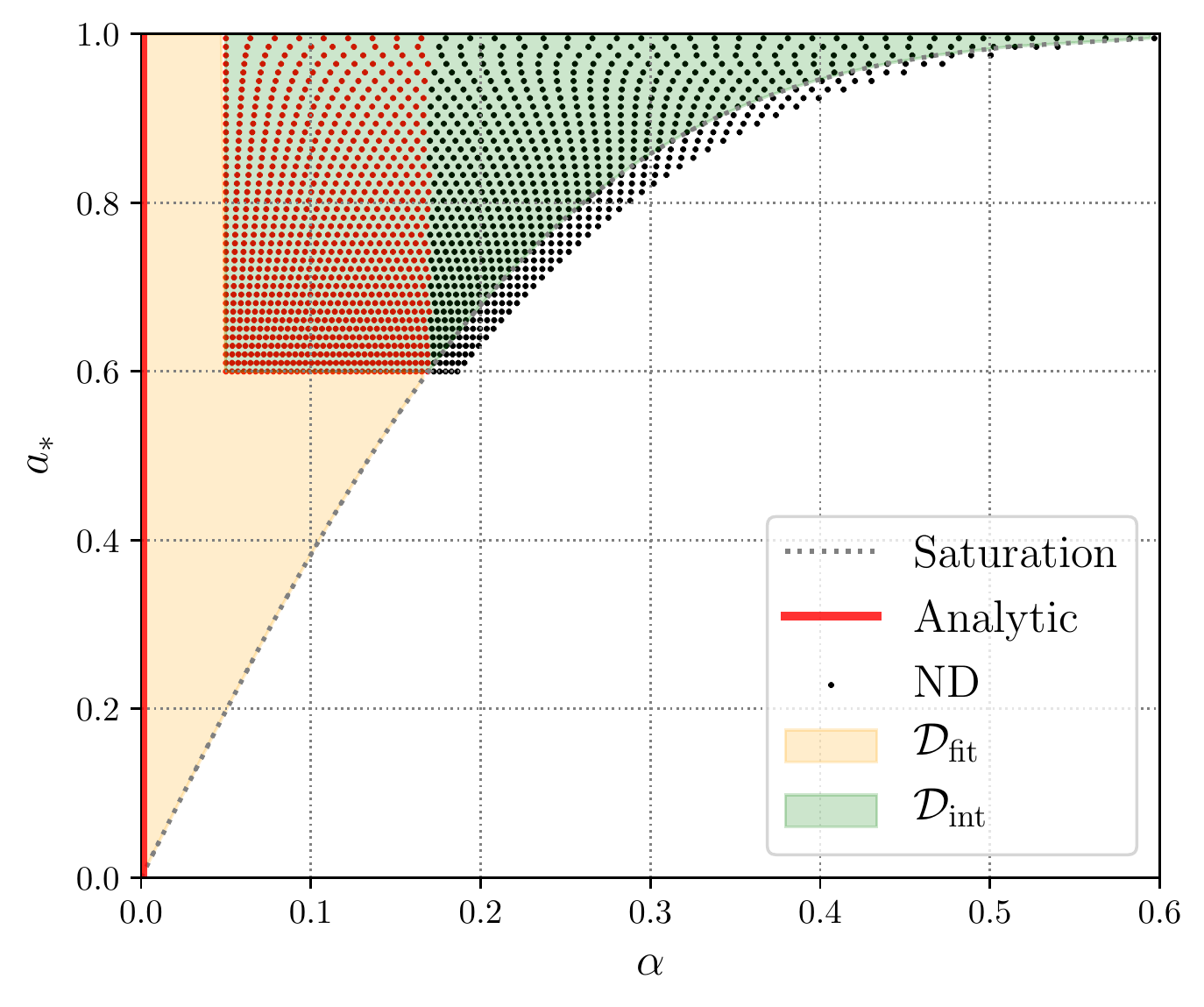}
\caption{The parameter space of the superradiant $m_V=1$ vector mode.  It is
made up of relativistic regime $\mathcal{D}_{\rm int}$, where \texttt{SuperRad} employs
interpolation functions based on the numerical data (labelled ND) to determine
a given quantity $q(\alpha,a_*)$ and a lower $\alpha$ region $\mathcal{D}_{\rm fit}$,
where numerical calibration is necessary to augment the expressions valid in
the Newtonian limit $\alpha \to 0$ (indicated by a red line). For illustration
purposes, we show only $40^2$ of the $320^2$ data points used in
\texttt{SuperRad}.  The gray dashed line marks the saturation point of the
superradiance instability, i.e., $\omega_R=\Omega_H(a_*)$. In this case, the
red data points are used for calibration in $\mathcal{D}_{\rm fit}$.}
\label{fig:pspace}
\end{figure}

In \figurename{ \ref{fig:pspace}},
we show the parameter space for the $m_V=1$ massive vector as an illustrative
example. For a given quantity $q(\alpha,a_*)\in \{\omega_R,\omega_I,\partial_t
\omega_R\}$ (in units of $M$), we numerically calculate its value in the 
relativistic regime, but for computational reasons, do not extend our calculations 
deep into the small-$\alpha$ regime. We want to match on to 
analytic results $q_N$ that are valid only in the Newtonian limit, when $\alpha \ll 1$.
We do this by dividing the parameter space in $(\alpha,a_*)$ into two regions.
In the relativistic regime, labelled $\mathcal{D}_{\rm int}$, 
we merely interpolate between the numerically computed points using the 
interpolation polynomial $I_R(\alpha,a_*)$. 
In the regime where $\alpha$ is smaller, labelled $\mathcal{D}_{\rm fit}$, 
we use a subset of the numerical results in $\mathcal{D}_{\rm int}$ (corresponding to the red points in Fig.~\ref{fig:pspace}) 
and fit the difference between these results and the analytic ones in a way that is guaranteed to recover the latter 
at sufficiently small $\alpha$. That is, we let
\begin{align}
q(\alpha,a_*)=\begin{cases} 
             q_N(\alpha,a_*)+g(\alpha,a_*), & (\alpha,a_*)\in \mathcal{D}_{\rm fit}, \\ 
             I_R(\alpha,a_*), & (\alpha,a_*)\in \mathcal{D}_{\rm int}. \end{cases}
\label{eq:quantity}
\end{align}
where $g$ is a fitting function chosen to give $q_N(\alpha,a_*)+g(\alpha,a_*) \to q_N(\alpha,a_*)$ as $\alpha \to 0$.
The specific choices of $\mathcal{D}_{\rm fit}$ and $\mathcal{D}_{\rm int}$ depend on the field
and azimuthal mode in question, and are determined by the accuracy of the
underlying methods (these are defined in App.~\ref{app:uncertainties}).
Note also that we are only interested in the part of the parameter
space where $\omega_R\leq m_{\sigma}\Omega_H$, since outside this range the
cloud will be exponentially decaying through absorption by the BH. 

In the relativistic part of the parameter space $\mathcal{D}_{\rm int}$, a set of $320^2$ waveforms are
generated for the azimuthal modes $m_\sigma=1$ and $2$, and for both the scalar
and the vector case. The grid of waveforms is uniformly spaced in the
coordinates $(y,a_*)$, with $y\in[0,1]$, defined by 
\begin{align}
y=\frac{\alpha-\alpha_0}{\alpha_1-\alpha_0}, 
\end{align} 
where $\alpha_0^{m_\sigma=1}=0.05$ and $\alpha_0^{m_\sigma=2}=0.25$, 
while $\alpha_1$ is the solution to 
\begin{align} 
\beta \alpha_1\left[1-\frac{\alpha_1^2}{2n_\sigma^2}\right]=m_\sigma M\Omega_H(a_*),
\end{align} 
with $\beta=0.9$, and $n_\sigma$ is the cloud's principle number defined below in
\eqref{eq:principlequantumnumbers}. 
This choice of $\alpha_1$ is made so as to guarantee that $y=1$ lies outside the superradiant
regime, and thus that the saturated state $\omega_R=m_{\sigma} \Omega_H$ lies within the grid. The boundary $y=1$ corresponds to the large-$\alpha$ boundary of the numerical data in \figurename{ \ref{fig:pspace}}, beyond the superradiant saturation.

\subsection{Oscillation frequencies} \label{sec:frequency}

The real part of the superradiantly unstable field's frequency
determines the cloud's oscillation about the BH
\begin{align}
A_\mu,\Phi\sim \cos( \omega_R t),
\end{align}
and also sets the characteristic GW frequency $f_{\rm GW}=\omega_R/\pi$ (up to self-gravity corrections). 
Because of the BH's gravitational potential, a bound massive
particle has a frequency $\omega_R<\mu$. Expanding
\eqref{eq:fieldeq} to leading order in $\alpha$ yields a Schrödinger-type
equation with potential $U\sim \alpha/r$, at a radius $r$ away from the BH. In
this regime, the solutions are simple hydrogen-like bound states for scalar and
vector fields \cite{Detweiler:1980uk,Pani:2012bp}. The scalar states
are characterized by their angular momentum quantum number $\ell_S$, as well as
azimuthal mode number $-\ell_S\leq m_S\leq \ell_S$ and radial node number
$\hat{n}_S\geq 0$, while the vector states are identified by an analogous
definition of radial node number $\hat{n}_V\geq 0$, angular momentum number
$\ell_V$ and azimuthal index $m_V$, in addition to the polarization state
$\hat{S}\in\{-1,0,1\}$. With this, the oscillation frequencies of the scalar
and vector clouds are, in the non-relativistic limit,
\begin{align}
\omega_R = \mu\left(1-\frac{\alpha^2}{2n^2_\sigma}+C_\sigma[\alpha]\right),
\label{eq:freqnonrel}
\end{align}
where $C_\sigma[\alpha]$ includes higher order corrections.
In particular, we include terms of up to $\mathcal{O}(\alpha^5)$,
obtained by keeping sub-leading contributions in $\alpha$ when solving
\eqref{eq:fieldeq} \cite{Baumann:2019eav}, with the full expressions
for $C_\sigma$ given in appendix~\ref{app:fieldsolutions} [in particular
\eqref{eq:freqrelcorrectionsscalar} and \eqref{eq:freqrelcorrectionsvector}].
The state label $n_\sigma$, depends on the intrinsic spin of the field and is given by
\begin{align}
n_S=\ell_S +1 +\hat{n}_S, & & n_V=m_V+\hat{n}_V+\hat{S}+1.
\label{eq:principlequantumnumbers}
\end{align}
Notice, in the case of the scalar field, we follow the conventions of
Ref.~\cite{Baumann:2019eav}, while in the vector case, we follow
Ref.~\cite{Dolan:2018dqv}.  In the language of the previous section, the
expressions \eqref{eq:freqnonrel} are the Newtonian estimates
$q_N(\alpha,a_*)$.

We numerically estimate $\omega_R$ using the methods discussed in
appendix~\ref{app:fieldsolutions}, without assuming an expansion in small
$\alpha$. These estimates are calculated for $m_\sigma\in\{1,2\}$ for both
scalar and vector fields. Here, we simply summarize that our numerical methods
are more accurate and precise than the analytic estimates everywhere in
$\mathcal{D}_{\rm int}$. The waveform model provides accurate values for
$\omega_R$ in $\mathcal{D}_{\rm fit}$ using a fit to the numerical results. We
perform this fit using the ansatz
\begin{align}
\frac{\omega_R}{\mu}-1+\frac{\alpha^2}{2n^2_\sigma}-C_\sigma[\alpha]=\sum_{q,p}\alpha^p\hat{a}_{p,q}(1-a_*^2)^{q/2},
\label{eq:wrfit}
\end{align}
with appropriately chosen ranges for $p$ and $q$, to the numerical data in
a subset of $\mathcal{D}_{\rm int}$ (see appendix~\ref{app:uncertainties} for details). The right-hand side of \eqref{eq:wrfit}
corresponds to $g(\alpha,a_*)$, defined in the previous section. This ansatz
explicitly assumes the analytic estimates in the $\alpha\ll 1$ regime. Within
\texttt{SuperRad}, we combine all three of these ingredients as described in
\eqref{eq:quantity} to determine $\omega_R$ in the parameter space.  Therefore,
we ensure that \texttt{SuperRad} provides the most accurate and precise
estimate for frequencies of a given superradiant bosonic field around a fixed
Kerr BH background across the entire parameter space.  The correction of these
frequency estimates due to the self-gravity of the superradiant cloud is
discussed in Sec.~\ref{sec:freqshift}.

\begin{figure}[t]
\includegraphics[width=0.49\textwidth]{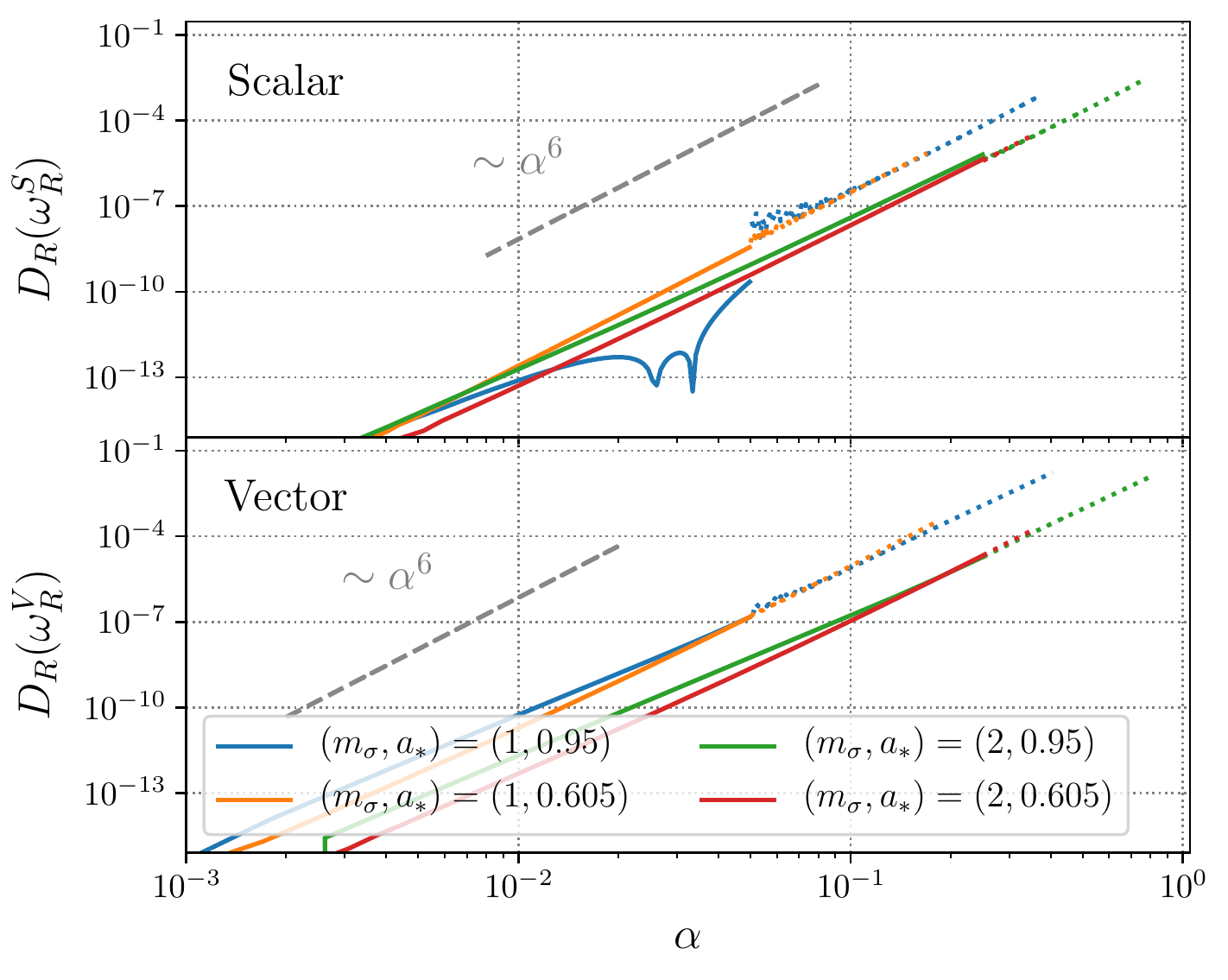}
\caption{The relative difference $D_R$, between the prediction for $\omega_R$
provided by \texttt{SuperRad}, and purely analytical non-relativistic estimates
given in \eqref{eq:freqnonrel} together with
\eqref{eq:freqrelcorrectionsscalar} and \eqref{eq:freqrelcorrectionsvector}.
Dotted lines indicate the $\mathcal{D}_{\rm int}$ region in \texttt{SuperRad}. We focus
on a few representative cases.}
\label{fig:frequenciesmodelerror}
\end{figure}

In \figurename{ \ref{fig:frequenciesmodelerror}}, we compare the available
analytic estimates, given in \eqref{eq:freqnonrel} [together with
\eqref{eq:freqrelcorrectionsscalar} and \eqref{eq:freqrelcorrectionsvector}],
with those provided by \texttt{SuperRad}.  As expected, the relative difference
between the analytic estimates and \texttt{SuperRad}'s decay as $\sim\alpha^6$
[the order of the leading-in-$\alpha$ unknown coefficient in the expansion of
\eqref{eq:freqnonrel}] in the Newtonian regime. For large spins $a_*$ and large
$\alpha$, i.e., in the relativistic regime, the analytic estimates have
relative errors up to $D_R(\omega_R)\lesssim 10^{-2}$.  In comparison to the
vector results, the analytic estimates for $\omega_R^S$ are
more accurate in the most relativistic regime.

\subsection{Instability timescales} \label{sec:timescale}

The imaginary part of the frequency $\omega_I$ sets the superradiant instability
timescale $\tau_I=1/\omega_I$ of the bosonic cloud,
\begin{align}
A_\mu, \Phi\sim e^{\omega_I t}.
\end{align}
In the non-relativistic limit $\alpha \to 0$, the cloud sits far away from the
BH and the flux across the horizon, and hence the instability growth rate,
tends towards zero. For small, but non-zero $\alpha$, the rates scale with a
characteristic power $\kappa$, i.e., $\omega_I M\sim\alpha^{\kappa}$. This
scaling depends on the type of field (scalar or vector) and the mode
considered. Furthermore, at saturation, i.e., when
$\omega_R=m_\sigma\Omega_H$, the ultralight particles cease extracting
rotational energy from the BH, such that the growth rate vanishes.
Combining these two limits, the general behavior of the instability growth rates
for both the scalar and the vector cases is
\begin{align}
\omega_I M=\alpha^{\kappa}(\omega_R-m_\sigma\Omega_H)2r_+G_\sigma(a_*,\alpha).
\label{eq:ratenonrel}
\end{align}
Here, $G_\sigma(a_*,\alpha)$ is a function of the BH spin, as well as $\alpha$,
and determines the leading order and sub-dominant-in-$\alpha$ contributions to
$\omega_I$. The scaling power $\kappa$, for scalar and vector fields are
\cite{Detweiler:1980uk,Pani:2012vp}
\begin{align}
\kappa_S=4m_S+5, & & \kappa_V=4m_V+2\hat{S}+5,
\end{align}
for the fastest growing configurations\footnote{Notice, in the relativistic
regime, it is non-trivial to identify the most unstable mode.}, and depend on
the azimuthal index $m_\sigma$ and the vector polarization state $\hat{S}$. The
leading order contributions in the scalar case~\cite{Detweiler:1980uk} 
and the vector case \cite{Rosa:2011my,Pani:2012bp,Baryakhtar:2017ngi,Baumann:2019eav} 
to $G_\sigma(a_*,\alpha)$ that we use
are given in Appendix~\ref{app:fieldsolutions} [in particular \eqref{eq:raterelcorrectionsscalar} and
\eqref{eq:raterelcorrectionsvector}, respectively]. 
These are Newtonian estimates that we use [$q_N(\alpha,a_*)$ in the language
of Sec.~\ref{sec:parameterspace}] for the imaginary frequency.

Similarly to the previous section, we utilize numerical techniques to obtain
accurate predictions for $\omega_I$ in the relativistic regime $\mathcal{D}_{\rm int}$
of the parameter space. The methods and their accuracy are outlined in
Appendix~\ref{app:fieldsolutions}. Here, we simply note again that the numerical
predictions are more accurate than the analytic Newtonian expressions everywhere in $\mathcal{D}_{\rm int}$. 
Similar to the
real part of the cloud's frequency, the analytic results obtained in the
Newtonian limit are connected with the numerical estimates in the $\alpha\sim
1$ regime by fitting\footnote{Notice a typo in eq. (A.2) of
\cite{Siemonsen:2019ebd}; it is fixed by $C_m\rightarrow 2C_m r_+$.} the ansatz
\begin{align}
\begin{aligned}
& \frac{\omega_IM \alpha^{-\kappa}G^{-1}_\sigma(a_*,\alpha)}{2r_+(\omega_R-m_\sigma\Omega_H)}-1\\
& \qquad =\sum_{p,q}\alpha^p\left[\hat{b}_{p,q}a_*^{q+1}+\hat{c}_{p,q}(1-a_*^2)^{q/2}\right],
\label{eq:wifit}
\end{aligned}
\end{align}
with appropriately chosen ranges for $p$ and $q$, to the numerical data obtained in $\mathcal{D}_{\rm int}$ (see Appendix~\ref{app:uncertainties} for details).
The right-hand-side of \eqref{eq:wifit} serves as
$g(\alpha,a_*)$ in the construction \eqref{eq:quantity} for $\omega_I$.
Analogously to the oscillation frequency, with this construction we ensure
\texttt{SuperRad} provides the most accurate and precise estimates for the
superradiance growth rate $\omega_I$ everywhere in the cloud's parameter space.

\begin{figure}[t]
\includegraphics[width=0.49\textwidth]{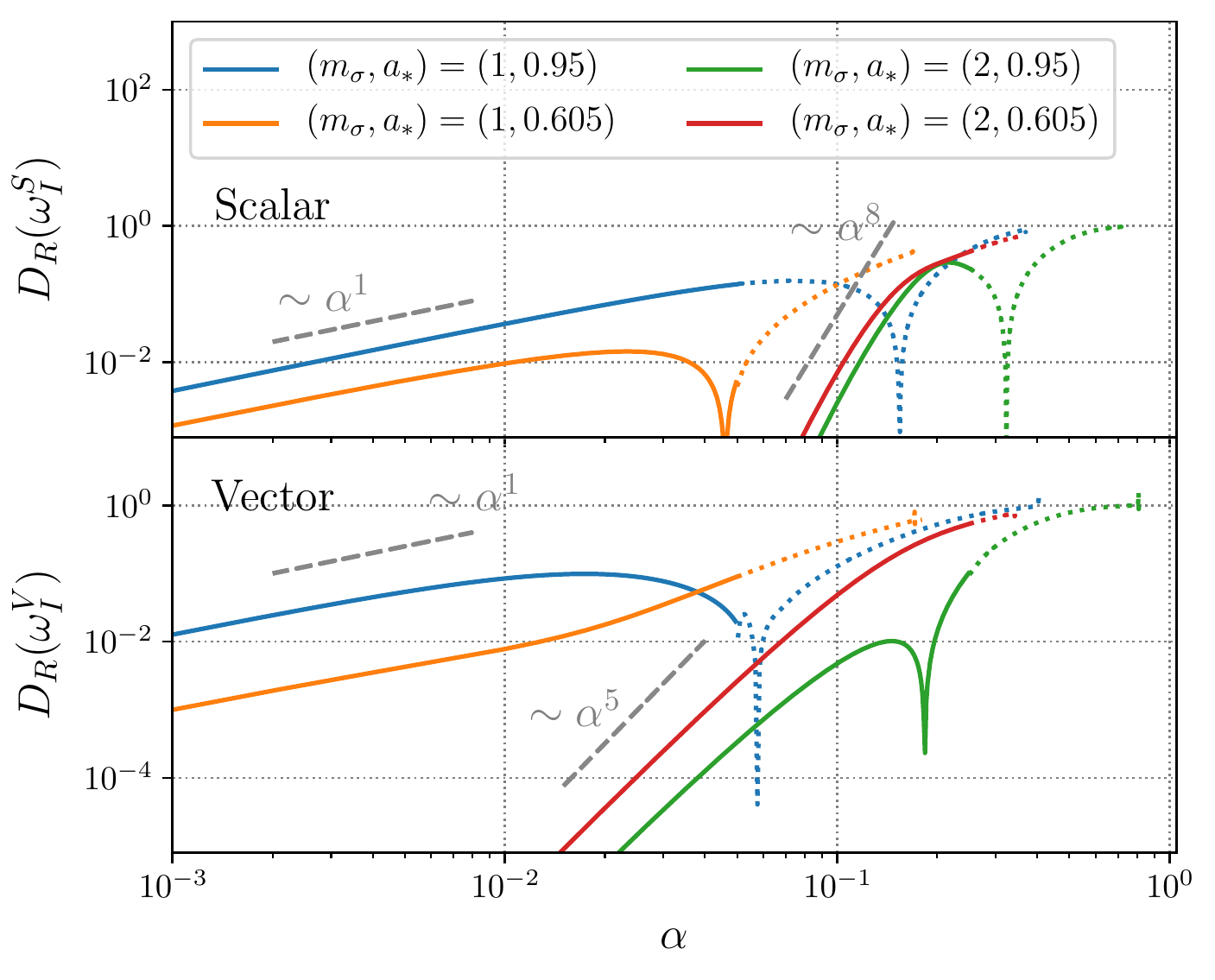}
\caption{The relative difference $D_R$ between the prediction for $\omega_I$
provided by \texttt{SuperRad}, and purely analytical non-relativistic estimates
given in \eqref{eq:ratenonrel} together with
\eqref{eq:raterelcorrectionsscalar} and \eqref{eq:raterelcorrectionsvector}.
Dashed lines indicate the $\mathcal{D}_{\rm int}$ region in \texttt{SuperRad}. We show the
same representative cases as in \figurename{
\ref{fig:frequenciesmodelerror}}.
} \label{fig:growthratesmodelerror}
\end{figure}

In \figurename{ \ref{fig:growthratesmodelerror}}, we illustrate the relative
differences between the analytic estimates using only \eqref{eq:ratenonrel}
together with \eqref{eq:raterelcorrectionsscalar} and
\eqref{eq:raterelcorrectionsvector}, and the estimates provided by
\texttt{SuperRad}. In the Newtonian regime, the relative difference approaches 
zero, while in the relativistic regime, the
relative error in the analytic estimates becomes $D_R(\omega_I)\sim\mathcal{O}(1)$
in both the scalar and the vector cases. Hence, using non-relativistic analytic
estimates in the relativistic regime can lead to large systematic
uncertainties in the instability rate. We indicate the leading-in-$\alpha$ scaling of the difference for each $m_\sigma$. An $\sim\alpha^1$-scaling is expected in principle for both $m_\sigma=1$ and $m_\sigma=2$, however, due to our choices of $p$ and $q$ in \eqref{eq:wifit} (see also Appendix~\ref{app:uncertainties}), the leading power is $>1$ for $\alpha\ll 1$ in the $m_\sigma=2$ case. For $\alpha\gtrsim 0.1$, the scaling decreases to the expected $\sim\alpha^1$.
 
\section{Frequency shift} \label{sec:freqshift}

So far, we have considered calculations that assume the bosonic field
can be treated as a test field on a Kerr background. Even for cases where the
boson cloud mass reaches $M_c\sim 0.1M$, treating the spacetime as Kerr, with
quasi-adiabtically changing parameters, gives a good
approximation to the nonlinear treatment~\cite{East:2017ovw,East:2018glu}.  However, in this section, we
address the effect of the self-gravity of the cloud, focusing in particular on
how it causes the characteristic increase in frequency of the cloud oscillation, and hence the frequency
of the emitted GW radiation. Though the cloud-mass induced
shift in the frequency is small, it will change as the cloud slowly dissipates
through gravitational radiation, affecting how long the GW
signal can be coherently integrated without taking this effect into account. 
Quantitatively estimating the contribution to the frequency from the finite cloud mass 
\begin{equation}
    \Delta\omega(M_c) = \omega(M_c)-\omega(M_c=0)
\end{equation}
(which we will assume to be real) is the subject of this section. We employ a
Newtonian approach, recovering and extending known results in the literature.
We then compare these results in the scalar case to a fully nonlinear approach
using synchronized complex fields around BHs.  

\subsection{Newtonian approach}
\label{ssec:freq_nr}
The Newtonian approach, utilized to estimate the cloud mass correction to the
frequency in Refs.~\cite{Baryakhtar:2017ngi,Isi:2018pzk,Baryakhtar:2020gao}, exploits the fact that in the
non-relativistic limit,
the energy density\footnote{In a spacetime, like Kerr, with asymptotically timelike Killing field $\xi^\mu$ and time-slice normal vector $n^\mu$, the energy density is defined as $\rho=n_\alpha\xi_\beta T^{\alpha\beta}$ through the scalar or vector field's energy-momentum tensor $T^{\alpha\beta}$.} $\rho$ is
spread out over large scales away from the BH, minimizing curvature effects. In
this limit, the cloud itself sources a Newtonian gravitational potential
$\Psi$, which follows the Poisson equation:
\begin{align}
\Delta_\text{flat} \Psi=4\pi \rho, & & \Psi(\textbf{r})=-\int d^3\textbf{r}' \frac{\rho(\textbf{r}')}{|\textbf{r}-\textbf{r}'|}.
\end{align}
Here, the coordinates $\textbf{r}$ can be
identified with spatial slices of Kerr, where gauge ambiguities disappear in the 
$\alpha\ll 1$ limit. Furthermore, while one might choose
$d^3\textbf{r}'=\sqrt{\gamma}d^3x'$, with the determinant of the
metric of a spatial slice of Kerr, a priori this is not more
consistent than simply setting $\gamma\rightarrow\gamma_\text{flat}$,
which is our choice. 
In this weak-field limit, the scalar wave equation \eqref{eq:fieldeq} is given by
\begin{equation}
\label{eom-eq}
    %(\omega - \mu_S) \phi(\textbf{r}) \approx -\frac{\nabla^2}{2\mu}\phi(\textbf{r}) + \mu (1 + g^{00})\phi(\textbf{r})
    (\omega - \mu_S) \Phi(\textbf{r}) \approx \left(-\frac{\nabla^2}{2\mu_S}-\frac{\mu_S M}{r}+\mu_S \Psi \right)\Phi(\textbf{r}) \ ,
\end{equation}
with $r=|\textbf{r}|$.
Taking the usual approximation that the shift in frequency at leading order in $\alpha$ is
given by evaluating the perturbed operator on the unperturbed eigenfunction, 
the self-gravity of a cloud with mass $M_c = \int d^3\textbf{r}\rho(\textbf{r})$ contributes a shift in frequency of
\begin{align}
\Delta \omega \frac{M_c}{\mu} &\approx \int d^3\textbf{r} \rho \Psi  \nonumber \\
&= 2\int  d^3 \textbf{r} \int_{|\textbf{r}'|<|\textbf{r}|} d^3 \textbf{r}' \frac{\rho(\textbf{r})\rho(\textbf{r}')}{|\textbf{r}-\textbf{r}'|} 
= 2 W.
\label{eqn:delta_omega}
\end{align}
We used the non-relativistic approximation that $\rho \approx  \mu^2 |\phi|^2$, and in the last line introduced
the total potential energy $W$ in the cloud.
We note that this factor of $2$ (from restricting the inner integral) is missing from some references~\cite{Baryakhtar:2017ngi,Isi:2018pzk}, but included in 
Ref.~\cite{Baryakhtar:2020gao}.
An equivalent derivation gives the same expression~\eqref{eqn:delta_omega} in the vector case as well.

We can further simplify the frequency shift calculation by considering a low multipole
approximation.  The denominator of~\eqref{eqn:delta_omega} can be expanded in
terms of spherical harmonics $Y_{\ell m}(\Omega)$, where $(\Omega) = (\theta,
\varphi)$ describes the angular dependence, as 
\begin{align}
\label{sph-harm-expansion-eq}
\frac{1}{|\textbf{r}-\textbf{r}'|}=\sum_{\ell=0}^\infty \sum_{m=-\ell}^\ell\frac{r'^\ell}{r^{\ell+1}} \frac{4\pi}{2\ell+1} Y_{\ell m}(\Omega)\bar{Y}_{\ell m}(\Omega'),
\end{align}
assuming $|\textbf{r}'|<|\textbf{r}|$. 
If we keep only the monopolar, i.e., the $\ell=0$, component of the density, which we 
can write in terms of the radial mass function $m_c(r)=\int d\Omega' \int_0^r dr' r'^2\rho(\textbf{r}')$,
then~\eqref{eqn:delta_omega} simplifies to
\begin{align}
\Delta\omega=- \frac{2\mu}{M_c}\int d^3\textbf{r}\frac{m_c(|\textbf{r}|)\rho(\textbf{r})}{|\textbf{r}|}.
\label{eqn:delta_omega_ss}
\end{align}

In general, there are non-vanishing higher order multipoles due to the non-trivial
azimuthal and polar dependencies of the cloud's energy densities that are neglected above. 
However, for the $m_S=1$, the error in the calculation of $W$ associated with making this 
monopole approximation, as opposed to considering higher multipole corrections, is $\approx2\%$ at leading order in $\alpha$ \footnote{At leading order in $\alpha$, only the quadrupole $\ell=2$ contributes non-trivially.}. 
In the $m_V=1$ case, all higher-order multipolar contributions are sub-leading in $\alpha$, since the Newtonian energy density is spherically symmetric. While the cloud states with larger azimuthal number have strong polar dependencies, the corrections from high-order multipoles is moderate. For the $m_V=2$ state, the quadrupolar contribution is $\approx2\%$ of the monopolar piece, at leading order in $\alpha$.
The frequency shift $\Delta \omega$ is then calculated for different modes
of the non-relativistic solutions to \eqref{eom-eq} for scalar clouds, and for 
corresponding non-relativistic vector cloud solutions.  
Expressions for $\Delta\omega$, valid for any azimuthal index $m_\sigma$, as well as 
a table listing the first few values, are given in \eqref{eq:deltaomegaexpressions} and in Table~\ref{tab:nonrel_shifts}, respectively.

\subsection{Quasi-relativistic}
\label{ssec:freq_qr}

These analytic expressions \eqref{eq:deltaomegaexpressions} are accurate in the Newtonian limit, i.e., $\alpha\ll 1$. Here, we extend the validity to the $\alpha\sim\mathcal{O}(1)$ regime, with the caveat that a more accurate nonlinear treatment, discussed in the next section, is ultimately necessary. Within \texttt{SuperRad}, we compute the frequency shift in the relativistic regime $\mathcal{D}_{\rm int}$ in a quasi-relativistic approximation, as in Ref.~\cite{Siemonsen:2019ebd}.
We take the relativistic field configurations (derived in Appendix~\ref{app:fieldsolutions}) in Boyer-Lindquist
coordinates and use them to compute the energy density $\rho$, which we then use
to compute the frequency shift $\Delta\omega$ using the monopolar Newtonian expression~\eqref{eqn:delta_omega_ss}. This approach explicitly assumes a linear dependence of the frequency shift on the cloud mass: $\Delta\omega\sim M_c$. Given these quasi-relativistic results in the relativistic regime of the parameter space, we follow the approach taken in Secs.~\ref{sec:frequency} and \ref{sec:timescale}, to calibrate a fit that assumes the analytic expressions \eqref{eq:deltaomegaexpressions} against the quasi-relativistic results in $\mathcal{D}_{\rm int}$. The fit ansatz is
\begin{align}
	\frac{M^2 \Delta\omega}{-\alpha^3 M_c}+F_\sigma=\sum_{p\geq 1} \alpha^p \hat{d}^\sigma_{p},
\label{eq:deltaomegafit}
\end{align}
where $F_\sigma$ contains the leading-in-$\alpha$ contribution, computed above and explicitly given in Appendix~\ref{app:frequencyshift}.

\begin{figure}
\includegraphics[width=0.49\textwidth]{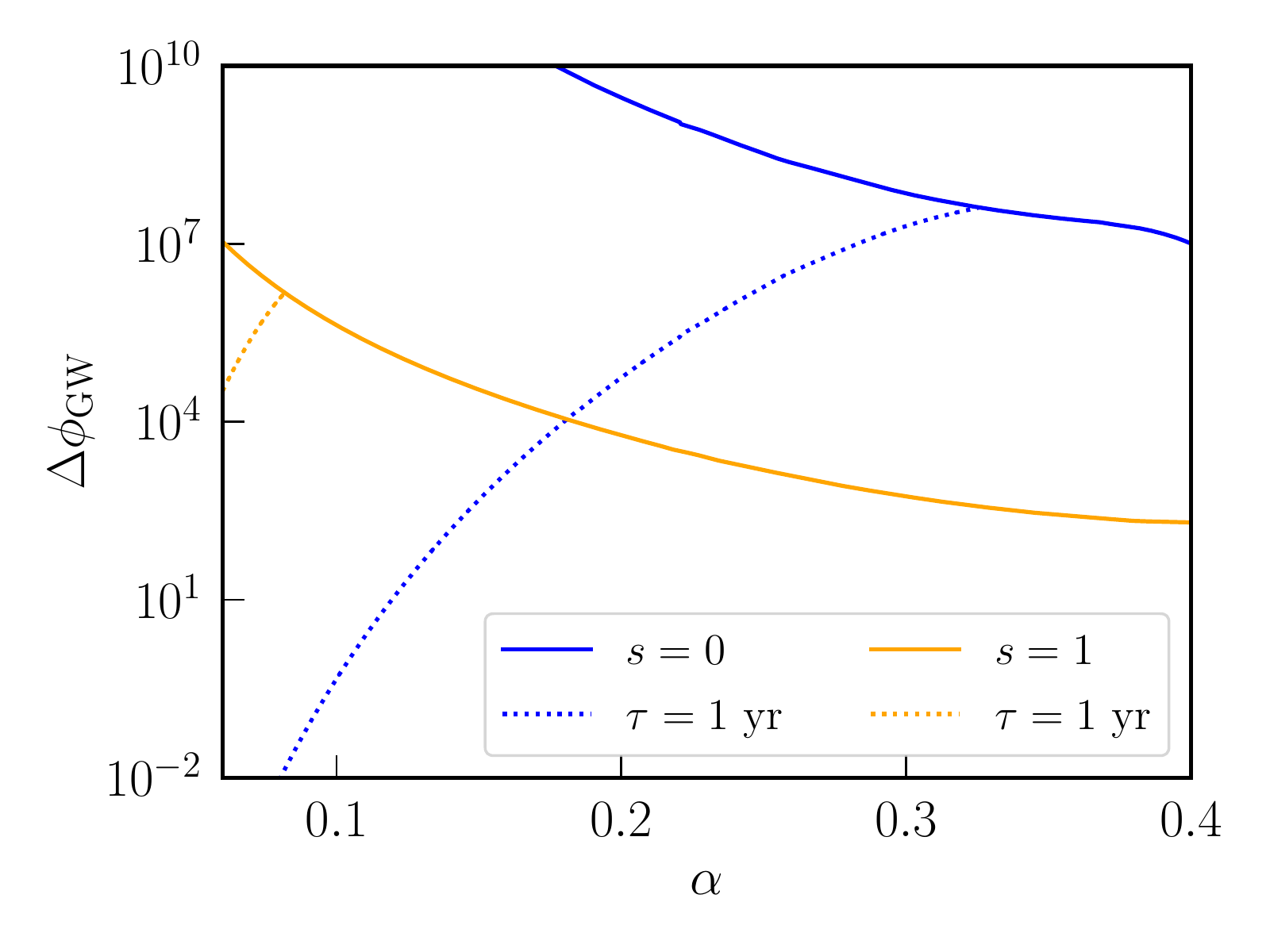}
\caption{
    The additional accumulated GW phase $\Delta \phi_{\rm GW}$ due to the
    increase in frequency as the boson cloud mass decreases [defined in
    \eqref{eqn:deltaphi}] for scalar (blue curves) and vector (orange curves) bosons. This phase is
    calculated beginning from when the cloud mass is maximum for a duration of
    $\tau_{\rm GW}$ (solid curves) and for one year (when $\tau_{\rm GW}>1$ yr;
    dotted curves).  We assume a BH with $M=50 \ M_{\odot}$ and $a_*=0.99$.
}
\label{fig:delta_phi}
\end{figure}

As a figure of merit for comparing how relevant this will
be in GW observations of boson clouds, we can calculate the extra
accumulated phase shift due to the frequency drift, 
using that $\omega_{\rm GW}=2\omega_R$,
\begin{equation}
    \Delta \phi_{\rm GW} = 2 \int_{t_{\rm max}}^{t_{\rm max}+\tau}\left[ \omega_R(t)-\omega_R(t_{\rm max})\right]dt,
    \label{eqn:deltaphi}
\end{equation}
where $t_{\rm max}$ is the time the cloud mass is at its maximum. 
We show this for the scalar and vector case in Fig.~\ref{fig:delta_phi}, taking 
the total time $\tau=\min(\tau_{\rm GW},1 \text{yr})$ to be either the characteristic time over which the GW signal decays $\tau_{\rm GW}$,
or one year, when $\tau_{\rm GW}> 1$ yr (assuming a $50 \ M_{\odot}$ BH). 
From the figure, we can see that $\Delta \phi_{\rm GW} \gg 1$ across the parameter 
space, except
for the scalar case when $\alpha \lesssim 0.1$.
Thus, properly accounting for this frequency shift is important to be
able to coherently integrate the GW signal. The diverging behavior of the $\tau=\tau_{\rm GW}$ curves in Fig.~\ref{fig:delta_phi} at low $\alpha$ is due to the steeper $\alpha$-scaling of the GW timescales compared with the frequency shift's scaling.

\subsection{Comparison to fully relativistic approach}
\label{ssec:freq_r}
\begin{figure}[t]
\centering
\includegraphics[width=0.47\textwidth]{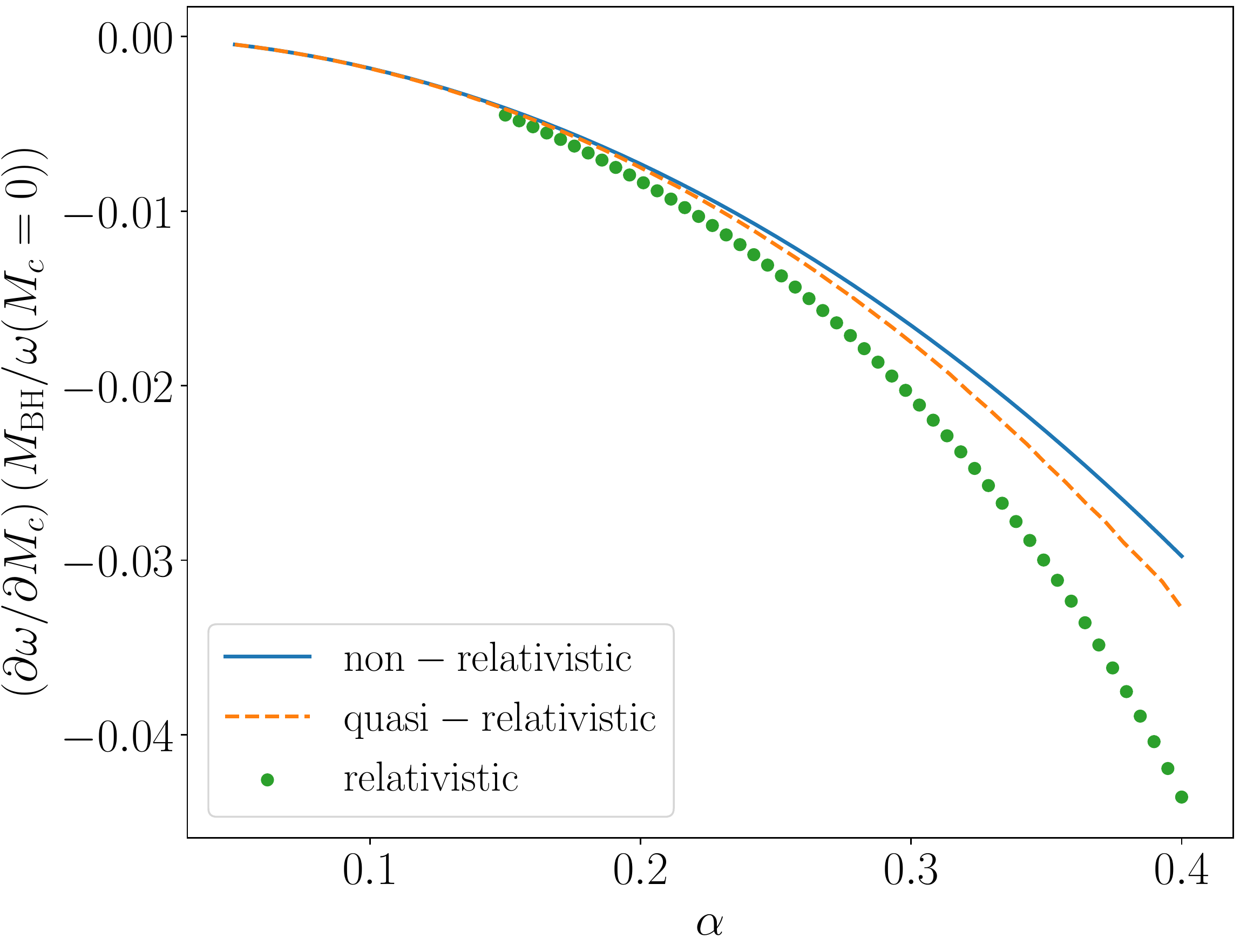}
\caption{A comparison of different approximations of the frequency shift due to
the boson cloud's self-gravity for a scalar field with $m_S=1$.  We compare the
non-relativistic (see Sec.~\ref{ssec:freq_nr}) and quasi-relativistic
(see Sec.~\ref{ssec:freq_qr}) approximations to the (leading order in
$M_c$ part) fully relativistic (labelled ``relativistic") relative frequency shift.  In
particular, we show, for fixed $\alpha$, $(\partial \omega/\partial
M_c)(M_c=0)\approx \Delta \omega/M_c$, where the equality is exact for the
non-relativistic and quasi-relativistic approximations.
\label{error-shifts}
}
\end{figure}

To gauge the error in the quasi-relativistic frequency shifts described above,
we compare them to numerically constructed, fully relativistic solutions.
Following Herdeiro \& Radu~\cite{Herdeiro:2015hr}, we construct stationary and
axisymmetric spacetime solutions to the full Einstein-Klein-Gordon field
equations consisting of a massive complex scalar field cloud with $\Phi\sim
e^{i m_S \phi-i\omega_R t}$ around a BH, satisfying the synchronization
condition $\omega_R = m_S \Omega_H$.  These can be thought of as oscillation
(or, equivalently, azimuthal angle) averaged versions of the scalar cloud
solutions.  By calculating how the frequency of the solution changes with $M_c$
at fixed $M$ and $\alpha$, we can obtain a fully-relativistic estimate for the
frequency shift $\Delta \omega$. The frequency shift is the part of the real frequency 
that is dependent on the boson cloud mass, $ \omega(M_c) = \omega(M_c = 0) + \Delta \omega(M_c)$.
For the values of cloud mass relevant to
superradiance, $\Delta \omega$ is, to a good approximation 
linear in $M_c$, as expected from the non-relativistic
results above. Therefore, here we compute a numerical estimate of
 $\partial \omega /\partial M_c$ at $M_c=0$ and fixed $\alpha$
(which is $\approx \Delta \omega /M_c$, to within $\sim 1\%$ for $M_c<0.04M$).
In Fig.~\ref{error-shifts}, we show how this compares, for $m_S=1$, to the
non-relativistic and quasi-relativistic results for the frequency shift.  From
there it can be seen that the quasi-relativistic estimate used by
\texttt{SuperRad} is slightly more accurate than the non-relativistic
expressions, but still noticeably underestimates the frequency decrease, by
$\approx 32\%$, for $\alpha = 0.4$. For small $\alpha$, all three calculations
give similar results, as expected. In particular, for $\alpha<0.15$, the difference 
in the quasi- versus the fully relativistic calculation is $<7\%$. 

We plan to include the fully relativistic frequency corrections in a future
version of \texttt{SuperRad}, and we defer details on constructing the
BH-complex boson cloud solutions, as well as the massive vector case (where
we expect comparable, if somewhat larger, relativistic corrections) to upcoming
work \cite{future_paper}.
We note that with such relativistic solutions, there is still theoretical error associated with taking a 
complex instead of real field (and hence axisymmetric spacetime).
However, we can estimate this by comparing $\Delta \omega$ calculated from~\eqref{eqn:delta_omega} 
using the axisymmetric energy density calculated from the complex scalar field
solution, to the same quantity calculated from just taking the real part, scaled to give the same energy
$\Phi \to \sqrt{2} \text{Re}[\Phi]$. 
We find the relative difference to be $5\times 10^{-5}$, indicating the theoretical error in
the frequency shift should be $<0.01\%$ for these relativistic
results.

\section{Gravitational waves}
\label{sec:gw} 
In the previous sections, we focused primarily on the conservative sector,
neglecting GW dissipation from the system. In what follows, we outline the
computation of the GW strain from the oscillating boson cloud in the source
frame.  The general procedure is to consider superradiant solutions to the
field equations \eqref{eq:fieldeq} as sources for the linearized Einstein
equations. These source linear metric perturbations around the BH, which then
propagate on an (approximately) fixed Kerr spacetime towards the observer.
Analogous to the approach outlined in Sec.~\ref{sec:cloudproperties}, we use 
numerical calculations of the emitted GWs that are valid in the
relativistic regime, and combine those with input from analytic calculations
that are valid in the Newtonian regime, $\alpha\ll 1$, to cover the entire
parameter space.  In contrast, however, to the quantities calculated in
Sec.~\ref{sec:cloudproperties}, in several cases only the leading order scaling
of the GW power and strain with $\alpha$ is known, while the coefficient can be 
fixed accurately only with numerical methods.

In the following, we begin by outlining the conventions used in the literature
and in \texttt{SuperRad} in Sec.~\ref{sec:conventions}. We then discuss the
emitted GW energy flux and the polarization waveform, as well as the GW modes
in the source frame in Sec.~\ref{sec:gwpowerstrain}. 

\subsection{Conventions} \label{sec:conventions}
At a large distance $r$ away from the source, the
GWs in the source frame are captured by the polarization waveform
%$h=h_+-ih_\times$: 
\begin{align} 
h=h_+-ih_\times= \frac{\mathcal{A}}{r} e^{-i\phi_\text{GW}(t)}\psi(\theta)e^{im_{\rm GW} \varphi}.  
\end{align} 
The GW frequency is just twice the cloud oscillation frequency, hence
\begin{align} 
\phi_\text{GW}(t) = 2\int \omega_R(t) dt.
\end{align} 
As discussed in Sec.~\ref{sec:freqshift}, the frequency will change over
time as the cloud first grows exponentially, and then decays through GW
dissipation.  The azimuthal dependence is fixed exactly by that of the cloud in
question: $|m_{\rm GW}|=2m_\sigma$, whereas the polar contribution
$\psi(\theta)$ is dominated by the $\ell_{\rm GW}=m_{\rm GW}$
spin-($-2$)-weighted spherical harmonic mode, except in the relativistic regime
of the parameter space. The overall amplitude $\mathcal{A}$ of the signal
scales with a leading power in the gravitational fine structure constant of the
system: $\mathcal{A}\sim \alpha^q$. This amplitude is approximately independent
of BH spin and is proportional to the cloud's mass: $\mathcal{A}(t) \propto
M_c(t)$.

We decompose the polarization waveform $h$ into GW modes $h^{\ell m}$ with
${}_{-2}Y_{\ell m}(\theta,\varphi)={}_{-2}S_{\ell m}(\theta)e^{im\varphi}$, the
$-2$-weighted spherical harmonics\footnote{Normalized as $\int d\cos\theta \
{}_{-2}\bar{S}_{\ell m}(\theta) \ {}_{-2}S_{\ell m}(\theta) =
1$.}, leading to:
\begin{align}
h^{\ell m}=\int_{S^2}d\Omega h {}_{-2}\bar{Y}^{\ell m}(\theta, \varphi).
\label{eq:gwmodes}
\end{align}
Here, and in the following, we drop the subscripts ``GW" on the GW mode labels
$(\ell,m)$ for brevity, and distinguish these from the corresponding cloud
labels by referring to the latter with $(\ell_\sigma,m_\sigma)$. The
polarization waveform can be reconstructed as
\begin{align}
\begin{aligned}
h_+= & \ \frac{1}{r}\sum_{\ell\geq m}|h^{\ell m}|\left[ {}_{-2}S_{\ell m}+(-1)^\ell {}_{-2}S_{\ell -m} \right] \\
 & \qquad\times\cos(\phi_{\rm GW}+m\varphi+\tilde{\phi}_{\ell m}), \\
h_\times= & \ -\frac{1}{r}\sum_{\ell\geq m}|h^{\ell m}|\left[ {}_{-2}S_{\ell m}-(-1)^\ell {}_{-2}S_{\ell -m} \right] \\
 & \qquad\times\sin(\phi_{\rm GW}+m\varphi+\tilde{\phi}_{\ell m}),
\label{eq:polarizationwaveform}
\end{aligned}
\end{align}
where we used $h^{\ell, -m}=(-1)^\ell \bar{h}^{\ell m}$, and defined
$\tilde{\phi}_{\ell m}$ as the complex phase-offsets between different $h^{\ell
m}$. Finally, the total GW energy flux is
\begin{align}
P_{\rm GW}=\int d\Omega \frac{r^2(2\omega_R)^2|h|^2}{16\pi},
\end{align}
and can be decomposed into the power emitted in each polar GW $\ell$-mode as 
\begin{align}
P_{\rm GW}=P_{\rm GW}^{\ell=m}+P_{\rm GW}^{\ell=m+1}+P_{\rm GW}^{\ell=m+2}+\dots.
\label{eq:gwpower}
\end{align}
Due to the amplitude scaling $\mathcal{A}\propto M_c$, it is convenient to
factor out the dependence on the cloud's mass, and quote results only for the
rescaled GW power:
\begin{align}
\tilde{P}_{\rm GW}=P_{\rm GW} M^2/M_c^2.
\label{eq:rescaledgwpower}
\end{align}

\subsection{Gravitational wave power and strain} \label{sec:gwpowerstrain}

There are two main avenues to determine the strain $h$ in the context of BH superradiance.
On the one hand, there are frequency-domain approaches, solving a type of 
differential eigenvalue problem that assumes a BH background with linear
perturbations, while on the other hand, there are time-domain numerical
methods, which solve the full nonlinear Einstein equations. The former are readily
extended across the entire relevant parameter space, but do not capture nonlinear effects,
while the latter make no approximations, but carry relatively large
numerical uncertainties, and are not easily extended to cover large parts of the parameter
space. In this work, we mainly leverage frequency-domain methods, and validate
these against time-domain estimates, where applicable. These frequency-domain
methods can be classified into the ``flat" and the ``Schwarzschild"
approximations, as well as what we call the ``Teukolsky" approximation. The former
two are analytic estimates, valid only in the non-relativistic regime,
$\alpha\ll 1$, while the last named is a numerical approach,
which is computationally efficient only when $\alpha$ is not too small.
The details of these approximations are given in Appendix~\ref{app:gws}.
Ultimately, as done above, \texttt{SuperRad} combines the best of both worlds
and provides the most accurate estimates across the entire parameter space.

In the non-relativistic limit, the currently available results are of the form
\begin{align}
\tilde{P}_{\rm GW}=H \alpha^{\eta}.
\label{eq:nonrelPgw}
\end{align}
The respective $\alpha$ scalings for the GW power from scalar and vector superradiant clouds are \cite{Arvanitaki_precision,Yoshino:2013ofa,Brito:2014wla,Baryakhtar:2017ngi}
\begin{align}
\eta_S=4m_S+10, & & \eta_V=4m_V+6,
\end{align}
while the numerical coefficient $H$ depends on the type of approximation
employed. We quote all available results in Appendix~\ref{app:gws}, and focus
here solely on those associated with $m_\sigma=1$ cloud states. The
Schwarzschild approximation has been studied only in the $m_\sigma=1$ case,
resulting in \cite{Brito:2014wla,Baryakhtar:2017ngi}
\begin{align}
(H_S)_{\text{Schw.}}^{m_S=1}=\frac{484+9\pi^2}{23040}, & & (H_V)_{\text{Schw.}}^{m_V=1}=60.
\end{align}
These overestimate the true emitted GW power, while the ``flat" approximation \cite{Yoshino:2013ofa,Baryakhtar:2017ngi}
\begin{align}
(H_S)_{\text{flat}}^{m_S=1}=\frac{1}{640}, & & (H_V)_{\text{flat}}^{m_V=1}=\frac{32}{5},
\end{align}
is expected to underestimate the total energy flux. From comparing the Schwarzschild with the
flat approximation, it is clear that the non-relativistic approximations have
systematic uncertainties of roughly one order of magnitude.  Hence, even for
$\alpha\ll 1$, numerical techniques are required to reduce the uncertainty in
the coefficient $H$.

For this reason, and to extend the validity of the GW power and strain
predictions of \texttt{SuperRad} to the part of the parameter space with the
loudest signals, we utilize frequency-domain numerical techniques in the
Teukolsky approximation.  We outline the methods we use in
Appendix~\ref{app:gws}. Here, we simply state that our numerical results are
more accurate than either of the analytic approximation techniques, even for
moderately small $\alpha$.

As evident from \eqref{eq:nonrelPgw}, the GW emission is independent of the BH
spin $a_*$ in the Newtonian regime, while in the relativistic regime, the GWs
exhibit mild spin-dependence \cite{Yoshino:2013ofa,Siemonsen:2019ebd}. To simplify the
parameter space, we restrict to clouds in the saturated state; 
that is, we assume $\omega_R=m_\sigma \Omega_H$
\footnote{The
validity of this last condition is discussed below in
Sec.~\ref{sec:cloudevolution}.}, removing
the spin-dependence from the parameter space. As in the discussion in
Sec.~\ref{sec:cloudproperties}, there exists a relativistic regime,
$\tilde{\mathcal{D}}_{\rm int}$, in which accurate numerical predictions can be
obtained. For $\alpha\ll 1$, the function
\begin{align}
\tilde{P}_{\rm GW}=b\alpha^{\eta}+c\alpha^{\eta+1}+\dots,
\label{eq:nonrelPgwfit}
\end{align}
is used to fit against the numerical results. In general, $b\neq H_\sigma$;
that is, we fit even the leading order coefficient from the numerically
obtained Teukolsky estimates. However, we check explicitly that $(H_\sigma)_{\rm
flat}<b<(H_\sigma)_{\rm Schw.}$ for both the scalar and vector $m_\sigma=1$ 
cloud states in the $\alpha\ll 1$ regime. \texttt{SuperRad} employs cubic-order interpolation in
$\tilde{\mathcal{D}}_{\rm int}$, and uses fits of the type \eqref{eq:nonrelPgwfit} for
$\alpha\in\tilde{D}_{\rm fit}$. 

\begin{figure}[t]
\includegraphics[width=0.49\textwidth]{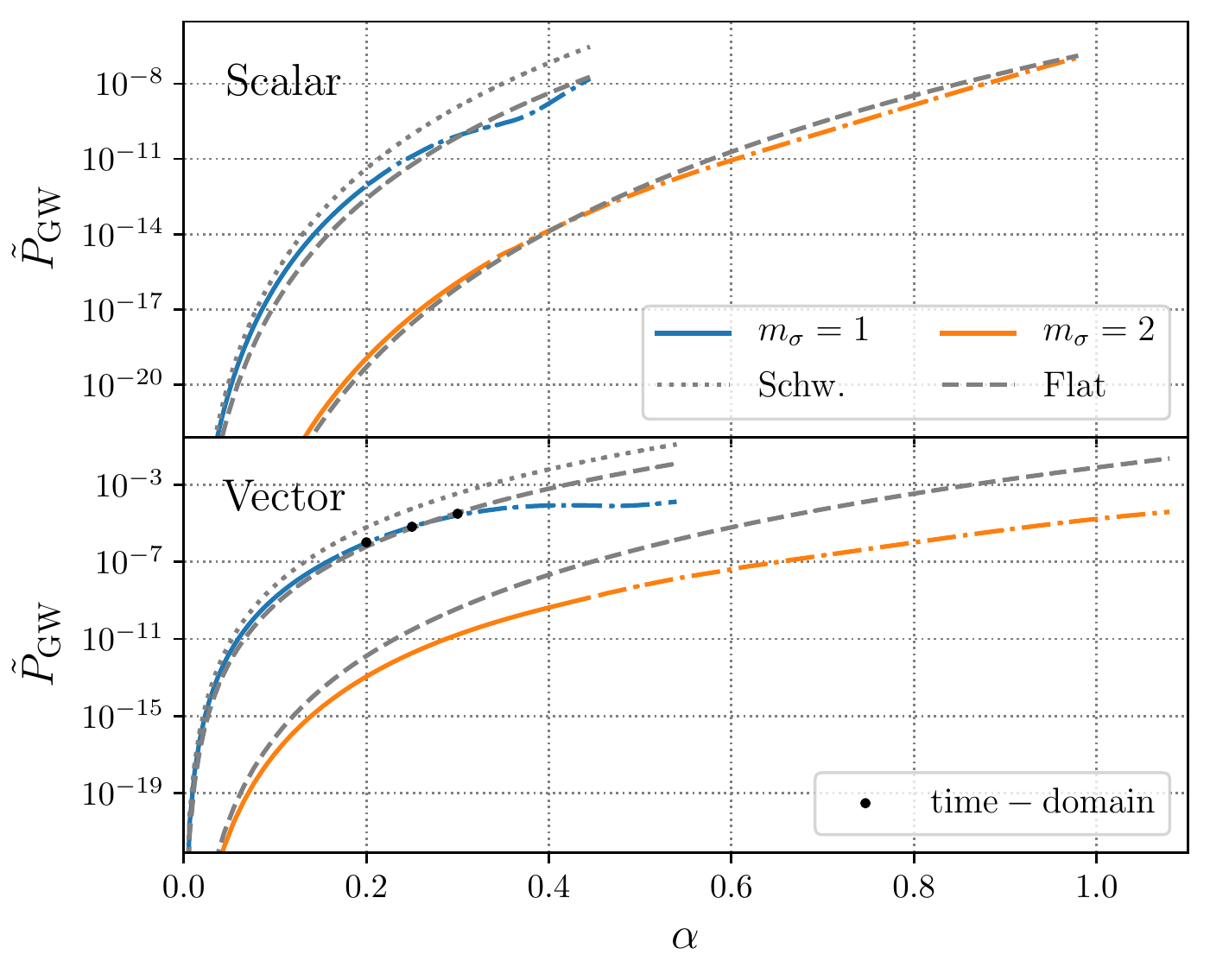}
\caption{
We show the mass-rescaled GW power $\tilde{P}_{\rm GW}$, defined in
\eqref{eq:rescaledgwpower}, emitted by the scalar and vector clouds with azimuthal number
$m_\sigma=1$ and 2 at the saturation point, $\omega_R=m_\sigma\Omega_H$,
comparing the Schwarzschild ``Schw." and the flat approximations to
\texttt{SuperRad} \textit{(colored lines)}, and time-domain estimates obtained
in \cite{East:2017mrj,East:2018glu}. Dash-dotted colored lines indicate where
\texttt{SuperRad} uses interpolation of numerical results over fits of the type \eqref{eq:nonrelPgwfit}.}
\label{fig:GWpower}
\end{figure}

In \figurename{ \ref{fig:GWpower}}, we compare the various calculation of the
the GW power to the predictions by \texttt{SuperRad}.  In the Newtonian limit,
\texttt{SuperRad} differs from \eqref{eq:nonrelPgw} due to the fit
\eqref{eq:nonrelPgwfit}, allowing different leading-$\alpha$ coefficients. The
underlying numerical results are more accurate (see Appendix~\ref{app:gws} for
details), allowing us to conclude that the estimates provided by
\texttt{SuperRad}, are more accurate than the Schwarzschild or flat
approximations. The analytic estimates for $\tilde{P}_{\rm GW}$ are worse for
$m_\sigma=2$; we use those results only to inform the leading-$\alpha$ scaling behavior. We also show time-domain results from evolving the full
nonlinear Einstein-Proca equations~\cite{East:2017mrj,East:2018glu} for a few
points. These agree with the Teukolsky calculations to within the numerical
error of the simulations. 

\begin{figure}[t]
\includegraphics[width=0.49\textwidth]{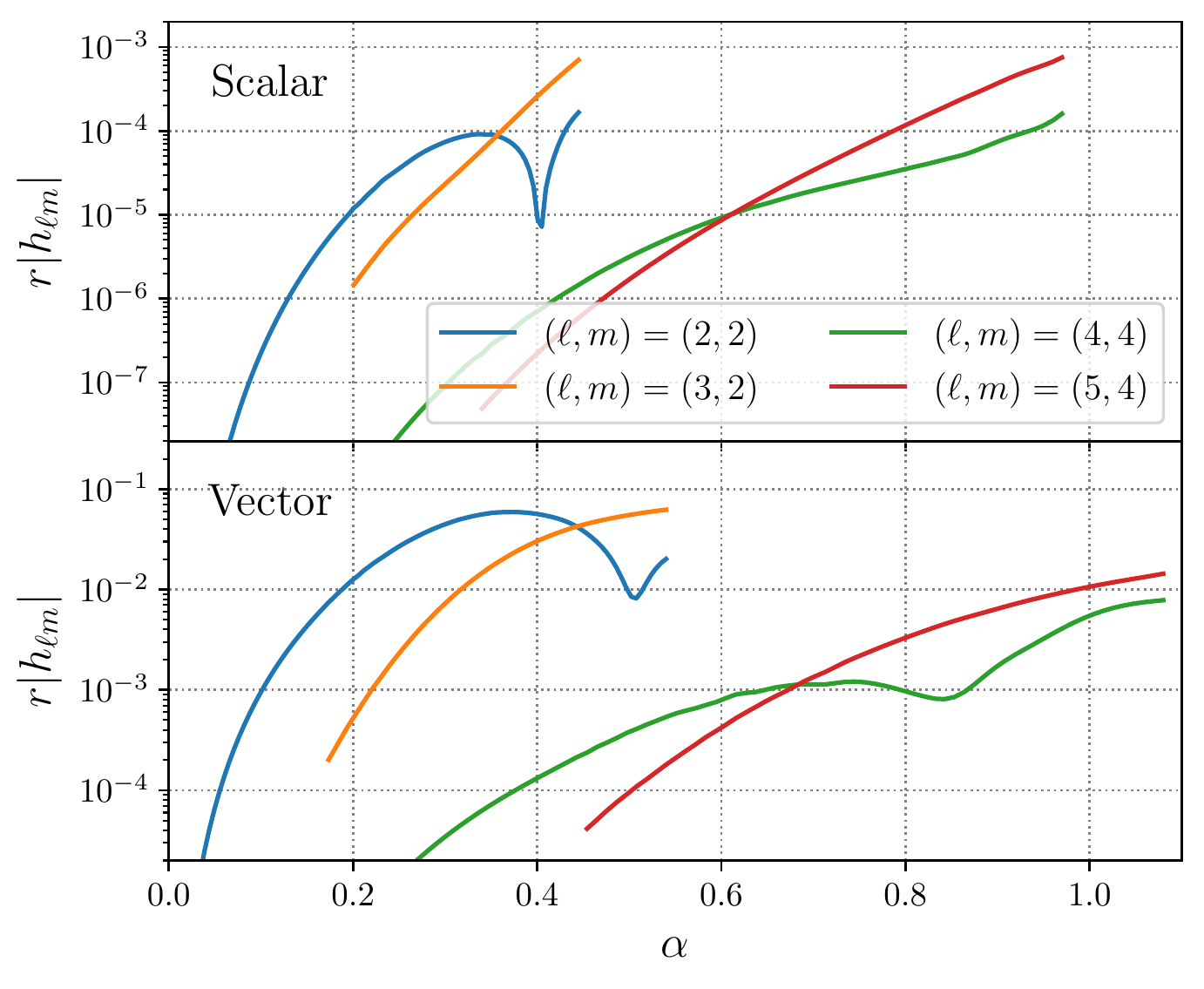}
\caption{
We show the magnitudes of the GW modes $h_{\ell m}$, defined in
\eqref{eq:gwmodes}, which are sourced by $m_\sigma=1$ and $2$ scalar and vector
boson clouds at saturation ($\omega_R=m_\sigma\Omega_H$) as functions of
$\alpha$. Notice that $\ell\geq 2m_\sigma$.}
\label{fig:GWstraim}
\end{figure}

In \figurename{ \ref{fig:GWstraim}}, we show the GW modes provided by
\texttt{SuperRad}, as defined in \eqref{eq:gwmodes}, over the entire parameter
space, assuming the saturation condition. As expected from the non-relativistic
results, the quadrupolar contribution $h_{22}$ dominates throughout most of
the parameter space, except in the most relativistic regime, where $h_{32}$
increases in importance (and equivalently for $h_{44}$ and $h_{54}$). This
behavior implies a constant phase shift between the two involved multipolar
components. Hence, there is an $\alpha$-range where $|h_{22}|\sim|h_{32}|$
(and $|h_{44}|\sim|h_{54}|$), which means that the phase difference
$\tilde{\phi}_{22}$ (and $\tilde{\phi}_{44}$), defined in
\eqref{eq:polarizationwaveform}, introduces a non-trivial phase-offset between
the two involved polar modes. 

\section{Growth and decay of boson cloud} \label{sec:cloudevolution}
In this section, we address how the superradiant instability and GW 
calculations can be combined to calculate the evolution of the boson cloud,
which determines the evolution of the amplitude and frequency of the GW signal.

A boson cloud around a spinning BH evolves as
the cloud extracts energy and angular momentum from the BH through the
superradiant instability.  During this process, the cloud also loses energy and
angular momentum to gravitational radiation. In a quasi-adiabatic
approximation, the evolution of this system is given by
\begin{align}
\begin{aligned}
\dot{M_c} & = 2\omega_I M_c+P_{\rm GW}, \\
\dot{M} & =-2\omega_I M_c, \\ 
\dot{J} & =-\frac{2 m_{\sigma} \omega_I}{\omega_R}M_c,
\label{eqn:cloud_evo}
\end{aligned}
\end{align}
where $\omega_R$, $\omega_I$, and $P_{\rm GW}$ are functions of the cloud mass
and BH mass and spin. The evolution of the boson cloud can be roughly divided
into two phases. In the first phase, the cloud grows exponentially, with the
mass going like $M_c\sim\exp(2\omega_I t)$, with the growth eventually saturating as
the BH is spun down and $\omega_I$ becomes small as
$m_{\sigma}\Omega_H$ decreases towards $\omega_R$. This is followed by the
gradual dissipation of the boson cloud through gravitational radiation. 
Since during this time $-\dot{M}_c\approx P_{\rm GW} \propto M_c^2$,
\begin{align}
M_c(t) \approx \frac{\bar{M}_c}{1+(t-t_{\rm max})/\tau_{\rm GW}}
\end{align}
where the cloud mass reaches a maximum $\bar{M}_c$ at $t=t_{\rm max}$
and $\tau_{\rm GW}:=\bar{M}_c/P_{\rm GW}$.

\begin{figure}[t]
\includegraphics[width=0.49\textwidth]{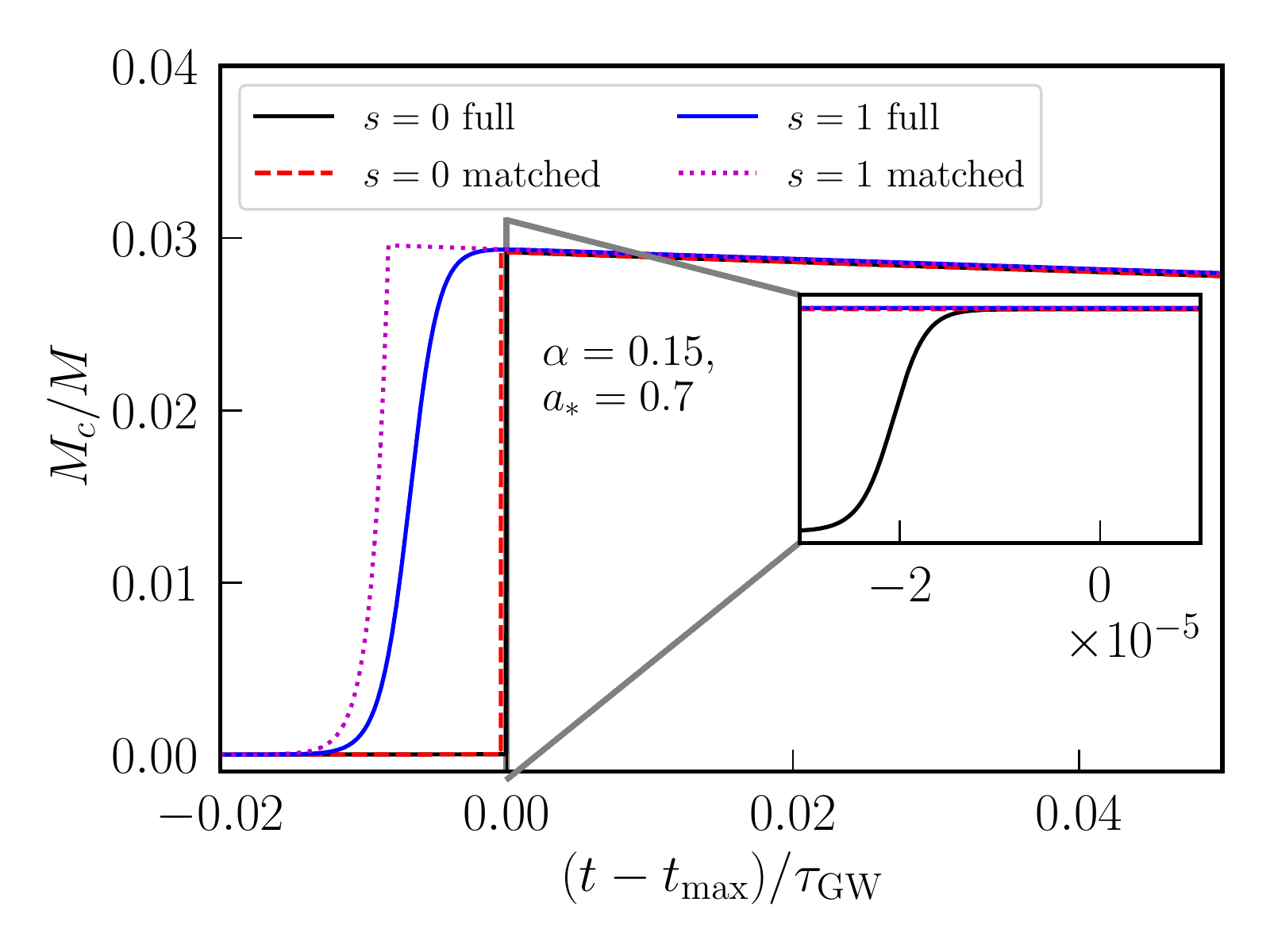}
\caption{ 
An example evolution of the boson cloud mass as a function of time for scalars
($s=0$) and vectors ($s=1$) with $\alpha=0.15$ and $a_{*}=0.7$.  The plot
compares the evolution determined by evolving the full
equations~\eqref{eqn:cloud_evo} (solid lines, labelled ``full"), to an
approximation that matches together constant exponential growth to GW-dominated
decay (dotted and dashed lines, labelled ``matched"). Time is normalized by the
gravitational dissipation timescale in either case, and the offset adjusted so
that the maximum value of $M_c$ occurs at zero for the full evolution cases,
and the matching value of $M_c$ is obtained for the corresponding matched
evolution cases. The inset shows a zoom in of the end of the exponential growth phase
for the scalar case (in particular the full evolution).
}
\label{fig:cloud_evo}
\end{figure}

In \figurename{ \ref{fig:cloud_evo}}, we plot an example of the evolution of the cloud
mass for both scalar and vector bosons. In both cases, $\tau_I \ll \tau_{\rm
GW}$, so that the exponential growth phase takes place on a much shorter time
scale than GW dissipation. However, the ratio
$\tau_I/\tau_{\rm GW}$ is markedly smaller in the scalar case compared to the
vector one. In addition to the full evolution of the cloud as determined by
\eqref{eqn:cloud_evo}, in \figurename{ \ref{fig:cloud_evo}} we also plot a simple
approximation where the maximum cloud mass is determined by solving for the BH
parameters where $\omega_R=m_\sigma \Omega_H$, and the evolution of $M_c$ after the
maximum is given solely by gravitational radiation, and the evolution of $M_c$
before the maximum is given by exponential growth with a fixed value $\omega_I$
given by the initial parameters.  

The \texttt{SuperRad} waveform model implements options for both the full cloud
evolution and the matched approximation. While the latter approximation is less
computationally expensive, as can be seen in \figurename{ \ref{fig:cloud_evo}},
it slightly overestimates the maximum cloud mass (by $\approx 0.04\%$ and
$0.8\%$, respectively, for the scalar and vector cases shown in the figure),
and underestimates the time for the cloud to reach its maximum. Thus, the more
accurate full cloud evolution is appropriate for scenarios when the signal
before the time when the cloud reaches saturation makes a non-negligible
contribution.  However, as noted above, our calculation of $\tilde{P}_{\rm GW}$
assumes $m_\sigma \Omega_H=\omega_R$, which is not strictly valid before the
saturation of the instability. Hence, there will be a discrepancy in the BH
spin used for the computation of the GW power.  This discrepancy is negligible
(i.e., below the numerical error of the methods, discussed in
Appendix~\ref{app:gws}) for $m_\sigma=2$, and for $m_\sigma=1$ assuming
$a_*<0.9$. It should be noted that this affects the GW emission before
saturation only, and also only systems with initial spin $a_*\gtrsim 0.9$. In
the vector $m_V=1$ case, the largest discrepancy occurs for $\alpha\approx
0.46$ and extremal spins, where the relative error from assuming the saturation
condition in the mass-rescaled GW power $\tilde{P}_{\rm GW}$ is $\approx 55\%$
(see Fig. 7 in \cite{Siemonsen:2019ebd}). For the scalar $m_S=1$ case this
discrepancy is at most $\approx 24\%$ around $\alpha\approx 0.36$. 

\section{LISA follow-up searches} \label{sec:lisafollowup}

The two main observational signatures of superradiant clouds, BH spin down and
GW emission, are sensitive to various systematic and statistical
uncertainties. Spin measurements have been used to exclude scalar and vector
mass ranges. Most of these constraints, however, rely on BH-spin estimates from
electromagnetic observations with significant systematic uncertainties. Spin
measurements of BHs in inspiraling binaries using GWs exhibit
large statistical uncertainties and make assumptions about the proceeding
history of the binary. Constraints from the stochastic GW background, assuming a population
of BH-cloud systems, rely on assumptions regarding the BH mass and spin population, 
in addition to position and distance uncertainties.
Lastly, searches for GWs from existing BHs observed in the electromagnetic
channel make assumptions about the past history of the observed BH, introducing
large systematic uncertainties. Clearly all of these methods rely on modeling
or assumptions with potentially substantial systematic uncertainties. 

One search strategy for GWs from superradiant clouds, however, evades these
assumptions: BH merger follow-up searches. These searches target BH remnants of
previously detected compact binary coalescences. The key advantages are the
knowledge of the complete past history of the targeted BH, as well as 
measurements of sky-position, spin, mass, and distance. Given these quantities,
accurate predictions of the subsequent superradiance instability and GW
emission are possible, enabling a targeted search for the latter in the
days/weeks/years following the merger. This removes the assumptions affecting
other search strategies, reduces the uncertainties to those coming from
the merger GW signal measurement of the remnant, and those of the waveform model
(discussed in the case of \texttt{SuperRad} below), and enables one to put
confident constraints on relevant parts of the ultralight boson parameter space,
or potentially to make a confident discovery.

In the context of the current generation of ground-based GW detectors,
follow-up searches for GWs from scalar superradiant clouds are likely
infeasible due to the small strain amplitudes \cite{Isi:2018pzk}. 
On the other hand, because of their faster growth rates and orders of magnitude stronger signals,
vector boson clouds are ideal candidates for these types of searches
\cite{Jonesinprep}. At design sensitivity, the advanced LIGO~\cite{TheLIGOScientific:2014jea}, advanced
Virgo~\cite{TheVirgo:2014hva}, and KAGRA~\cite{kagra} observatories will in
principle be sensitive to systems out to $\sim 1$ Gpc at a typical remnant BH
spin of $a_*=0.7$ and masses of $M\sim 100M_\odot$
\cite{Chan:2022dkt,Jonesinprep}. Undertaking follow-up searches targeting BHs
falling into this parameter range could target vector boson
masses roughly in the range of $\mathcal{M}_V\in (1\times 10^{-11},1\times
10^{-13})$ eV [see eq.~\eqref{eq:alphadefinition}]. In a similar fashion, LISA
could be sensitive to GWs from vector boson clouds with boson masses in the
$\mathcal{M}_V< 10^{-15}$ eV regime, inaccessible by ground-based detectors.

\begin{figure}[t]
\includegraphics[width=0.49\textwidth]{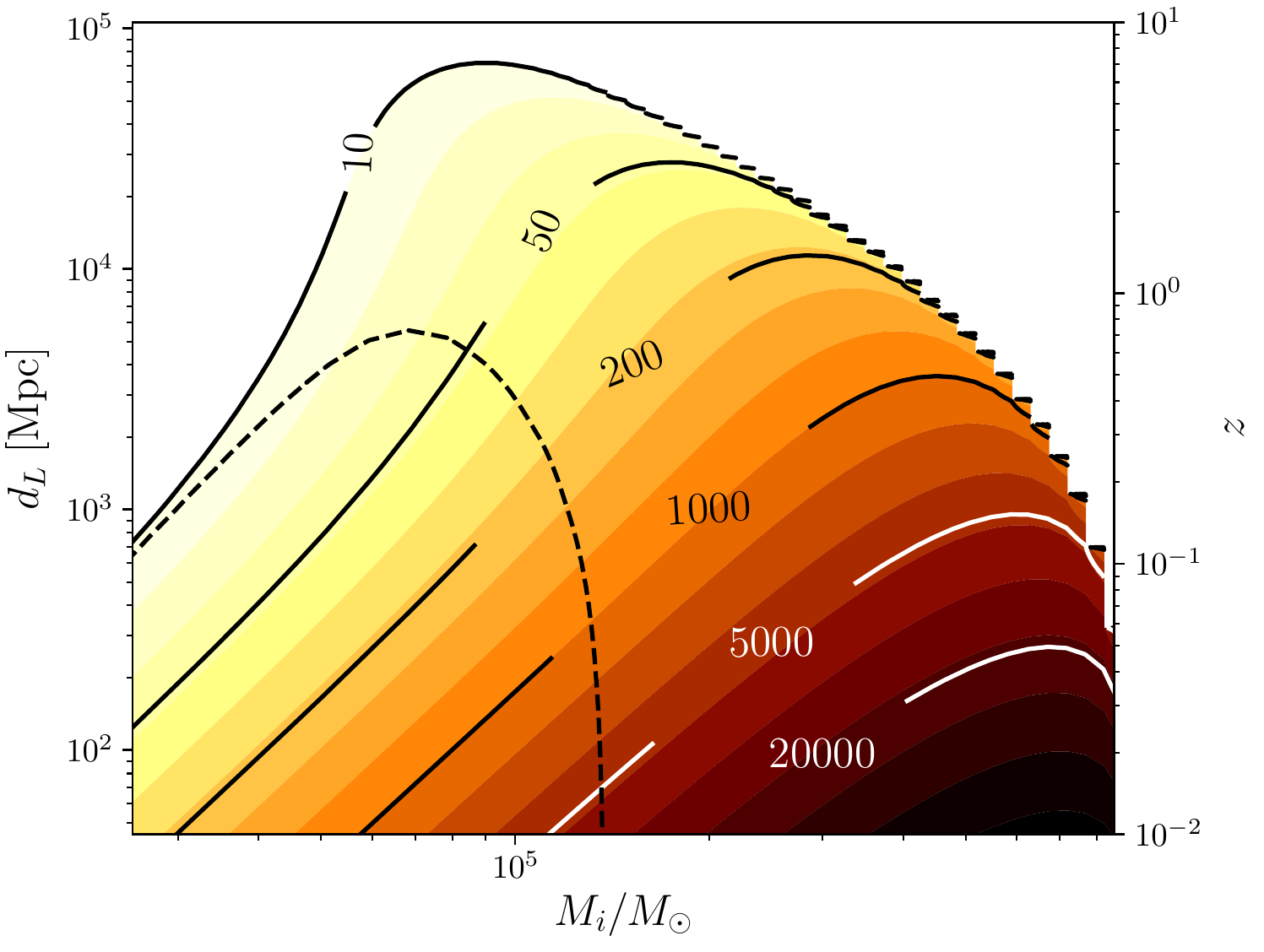}
\caption{We show the SNR \textit{(contour lines and color)} of GWs from vector
superradiant clouds around a fiducial BH of initial remnant source frame mass
of $M_i$ and spin $a_{*,i}=0.8$ as a function of luminosity distance $d_L$ and
redshift $z$, assuming a standard $\Lambda$CDM cosmology and $\alpha=0.2$. For
comparison, we also consider an initial spin of $a_{*,i}=0.7$ showing the $\rho_{\rm SNR}=10$ contour \textit{(dashed black line)}, assuming $\alpha=0.15$.}
\label{fig:lisafollowup}
\end{figure}

In the following, we analyze the prospects of follow-up searches for GWs from
vector superradiant clouds around supermassive binary BH merger remnants with
LISA. The fundamental assumption of follow-up searches is that a \textit{new}
superradiant cloud forms around the remnant after merger. If either of the
constituents already posses a superradiant cloud, it is expected to be depleted
before or during merger for nearly equal mass-ratio ($q\sim 1$) systems
\cite{Baumann:2018vus}. Even for $q>1$, depending on $\alpha$, clouds around
the constituents of the binary are efficiently removed before merger
\cite{Baumann:2018vus,Berti:2019wnn,Takahashi:2021yhy,Takahashi:2021eso}.
LISA is expected to see at
least a handful of such mergers over the mission lifetime of four years
\cite{Berti:2006ew,Micic:2007vd}. Therefore, to estimate the detection horizon,
we assume a fiducial supermassive binary BH merger remnant detection that occurs one year
into the mission. After merger at redshift $z$, residual ultralight vector densities around the
remnant, or quantum fluctuations, trigger the superradiance
instability\footnote{Notice, an equal-mass, non-spinning binary BH merger results in 
remnant BH with $a_*\approx 0.7$.} leading to the complete cloud formation,
and hence the peak of the GW signal, on
timescales of at most $t_c\approx \tau_I(1+z)\log(M_c/\mathcal{M}_V)/2$ in the detector frame. 
Over most of the
parameter space, these signals will last for longer than the remaining three
years of the LISA mission, leaving an observing time of $T_{\rm obs}=3-t_c$
years. We determine the maximum detection horizon of GWs from vector superradiant
clouds by considering the optimal signal-to-noise ratio (SNR) $\rho_{\rm SNR}$ with the
LISA sensitivity curve (details can be found in Appendix~\ref{app:snr}). Making
these assumptions, we illustrate the detection horizon of LISA for such events
in \figurename{ \ref{fig:lisafollowup}}.

From \figurename{ \ref{fig:lisafollowup}}, we conclude that parts of the vector
boson mass parameter space can be probed with idealized follow-up GW searches from
supermassive binary BH remnants. Even for moderate initial spins of
$a_{*,i}=0.7$, GWs can be detected up to $z\lesssim 0.8$, while for slightly
more favorable initial spins of $a_{*,i}=0.8$, the GW emission is observable out to
$z\lesssim 8$. The merger rate of massive BH binaries is expected to peak around $M\sim 10^6
M_\odot$ for equal mass ratio systems, $q\lesssim 1$, and at $z\approx 2$
\cite{Mazzolari:2022cho,Henriques:2014sga,2020MNRAS4954681I}. For initial BH
masses $M_i>10^6\ M_\odot$, the cloud formation timescales are larger than the
mission duration, $t_c>3$ years, leading to a drop in SNR. At high redshifts,
the sensitivity of LISA is primarily limited by the short effective observation
times in the detector frame. Larger BH masses (lower boson masses) can be
accessed only with larger initial spins, or significantly longer mission
durations. Consulting \eqref{eq:alphadefinition}, vector boson masses roughly
around $\mathcal{M}_V\in (1\times 10^{-16},6\times 10^{-16})$ eV are within reach of
these follow-up search strategies with LISA.

These prospects are subject to a few caveats. First, we determined the detection horizon and sensitivity of LISA to GW from vector clouds around remnant supermassive BHs using the optimal matched filter SNR. What fraction of this total available SNR could be recovered from the data by a realistic search algorithm is an open question, even for ground-based detectors \cite{Jonesinprep}. Secondly, the merger rate of massive BH binaries has large uncertainties. If the true merger rate were peaked at redshifts of $z>5$, a realistic follow-up search would require a very favorable initial BH spin $a_{*,i}>0.8$ to access a meaningful part of the vector boson parameter space directly, or an outlier event much closer. 

\section{Discussion}

We have introduced a new BH superradiance gravitational waveform model
called \texttt{SuperRad}. This provides the superradiance instability growth
timescale $\tau_I$, the cloud oscillation frequency $\omega_R$, 
the GW frequency $f_{\rm GW}(t)$ and strain $h_{\times/+}$ in
the source frame as a function of time, the GW power $P_{\rm
GW}$, and the evolution of the boson cloud. The \texttt{SuperRad} model makes
use of all available analytic and numerical estimates for these observables,
and calibrates analytic fits against the numerical data to extend the
applicability across the \textit{entire} parameter space of the $m=1$
and $2$ scalar and vector superradiant clouds. The waveform model
\texttt{SuperRad} can be used to inform and interpret the results of
GW searches for ultralight scalar and vector BH superradiance.
This includes both blind and targeted searches for resolved continuous wave signals,
as well as searches for a stochastic GW background from BH-boson cloud systems.
It can also be used when interpreting BH spin measurements using
GW or electromagnetic observations.
Importantly, \texttt{SuperRad} is accurate in the relativistic regime where the 
observable signals will be the strongest. 

As the ultralight boson cloud dissipates through gravitational radiation, there
is a small increase in the frequency of the GWs due to the
changing self-gravity contribution of the cloud.  As illustrated above, even
though this frequency drift is small, because of the large number of
GW cycles that make up a typical superradiance signal, not
properly accounting for it can lead to the signal model going out of phase in a
fraction of the observing time.  Fully including this second-order effect
within BH perturbation theory is challenging, and the results in
\texttt{SuperRad} for the frequency evolution of the GW signal
use non-relativistic approximations. By comparing these to fully-relativistic
numerical calculations for the scalar boson case, we found that the former
underestimates the value of $\dot{f}_{\rm GW}$ by $\sim 30\%$ for the most
relativistic (i.e. $\alpha \sim O(1)$) cases, though the differences are
smaller for more typical parameters. In future work, we plan to include the
fully-relativistic results for the cloud-mass contribution to the frequency for
both scalar and vector bosons in \texttt{SuperRad}.  Though, given the
stringent accuracy requirements imposed by the typical signal timescales (see Fig.~\ref{fig:delta_phi}), it is
likely that fully-coherent signal analysis techniques (e.g., match filtering)
will still not be feasible in much of the parameter space, better predictions
for the GW frequency evolution are nevertheless important in guiding
the application of semi-coherent techniques.

Furthermore, we investigated the viability of follow-up searches for GWs from
ultralight vector superradiant clouds with LISA targeting remnants of observed
massive binary BH mergers.  We found that these searches are confident probes
of the ultralight vector boson parameter space around $\mathcal{M}\in (1\times
10^{-16},6\times 10^{-16})$ eV. With current estimates of the merger rate of
massive BH binaries, LISA will be sensitive to GWs from vector boson clouds
around remnants of these mergers out to redshift $z\lesssim 8$ at mass-ratio
$q\lesssim 1$ and remnant black hole masses of roughly $M\in(6\times
10^4,2\times 10^5)M_\odot$. Our basic analysis leaves various questions
unanswered. We assumed the total available signal-to-noise ratio can be
recovered by a realistic search algorithm, which is an overestimate even in the
case of ground-based detectors \cite{Jonesinprep}. As well, a more detailed
study folding in massive black hole binary merger rates with superradiant cloud
growth timescales and emitted GW luminosities could provide an estimate for the
expected number and mass ranges of merger events where LISA would be sensitive
to the GW signal from an ultralight vector boson.

\begin{acknowledgments}
We would like to thank Dana Jones, Andrew Miller, and Ling Sun for insightful discussions and comments on this draft. The authors acknowledge financial support by the Natural Sciences and Engineering Research Council of Canada (NSERC). Research at Perimeter Institute is supported in part by the Government of Canada through the Department of Innovation, Science and Economic Development Canada and by the Province of Ontario through the Ministry of Economic Development, Job Creation and Trade. This research was undertaken thanks in part to funding from the Canada First Research Excellence Fund through the Arthur B. McDonald Canadian Astroparticle Physics Research Institute.
\end{acknowledgments}

\appendix
\section{LISA signal-to-noise ratio} \label{app:snr}

For a given an initial BH spin and mass, as well as ultralight boson mass,
\texttt{SuperRad} provides predictions for the GW strain
$h_{+/\times}(t,R,\theta,\phi)$ at time $t$, (luminosity) distance $r$, and
angles $(\theta,\varphi)$ in the source frame. In the case of LISA, the detector
response functions $\tilde{X}_{+/\times}(\Theta,\Phi,\psi,f)$ relate the GW
strain in the source frame to the strain in the detector. The latter depend on
the source's sky-position $(\Theta,\Phi)$, polarization $\psi$, and
frequency\footnote{We neglect the motion of LISA and the source with respect to each
other.} $f$. Hence, the GW amplitude in the detector, $\tilde{h}_{\rm
det.}(f)$, in the frequency domain is given by
\begin{align}
\tilde{h}_{\rm det.}(f)=\tilde{X}_+ \tilde{h}_+(f)+\tilde{X}_\times \tilde{h}_\times(f),
\end{align}
where $\tilde{h}_{+/\times}(f)$ are the Fourier-transforms of
$h_{+/\times}(t,R,\theta,\varphi)$. Let $\langle\dots\rangle$ be the
sky/polarization average over $\Theta$, $\Phi$, and $\psi$, and let $\mathcal{R}(f)$ be
the frequency-dependent transfer function defined by the sky/polarization
average of the detector response $\langle \tilde{h}_{\rm det.}^*\tilde{h}_{\rm
det.}\rangle=\mathcal{R}(f)[|\tilde{h}_\times(f)|^2+|\tilde{h}_+(f)|^2]$. Then
the SNR $\rho_{\rm SNR}$ is (see e.g.,
Refs.~\cite{Robson:2018ifk,Babak:2021mhe})
\begin{align}
\frac{\rho_{\rm SNR}^2}{4}=\int_0^\infty df \frac{\langle \tilde{h}_{\rm det.}^*\tilde{h}_{\rm det.}\rangle}{S_n(f)}=\int_0^\infty df \frac{|\tilde{h}_\times(f)|^2+|\tilde{h}_+(f)|^2}{S_h(f)},
\end{align}
where $S_n(f)$ is the noise power spectral density of LISA, and $S_h(f)=S_n(f)/\mathcal{R}(f)$ is the LISA sensitivity curve. For all estimates, we use the conservative six months confusion noise projections.

Since \texttt{SuperRad} provides the time domain GW strain in the source
frame, we add the appropriate redshift factors and use fast-Fourier-transform
algorithms to numerically transform into the frequency domain. This is feasible
for shorter signals considered in follow-up searches. However, it becomes
increasingly computationally expensive with longer signals at smaller
$\alpha$.

\section{Superradiant field solutions} \label{app:fieldsolutions}

In this appendix, we briefly summarize the numerical methods we use to obtain
the scalar and vector estimates for the oscillation frequencies $\omega_R$ and
instability growth rates $\omega_I$ discussed in Secs.~\ref{sec:frequency}
and~\ref{sec:timescale}, respectively.  We also provide bounds on the precision
of our methods, and comment on the resulting uncertainties.

\subsection{Scalar field}

The real massive scalar wave equation \eqref{eq:fieldeq} has been extensively
studied in the context of asymptotically flat BHs. On a Kerr
background of mass $M$ and spin parameter $a$, Detweiler~\cite{Detweiler:1980uk} first derived
expressions for the superradiance instability rates and oscillation frequencies.
These results were refined in
various other works, e.g., Refs.~\cite{Dolan:2007mj,Yoshino:2013ofa,Baumann:2019eav}.
In this subsection, $\omega$, $\ell$, $n$, and $m$ refer exclusively to the scalar
mode numbers, hence, we drop the subscripts used throughout the main text, for
brevity.

Generally, due to the backgrounds symmetries, the most convenient scalar field
ansatz is of the form $\Phi=\text{Re}[R_s(r)S_s(\theta)e^{-i(\omega
t-m\varphi)}]$. With this ansatz, the field equations separate into a pair of
polar and radial second-order ordinary differential equations. The polar
equation can be identified with the spheroidal harmonic equation of spin-weight
$s=0$ and spheriodicity $c^2=-k^2a^2$, with $k^2=\mu^2_S-\omega^2$; the
solutions to this equation is the set of spheroidal harmonics,
${}_s\tilde{S}_{\ell m}(\theta;c)$, of spin-weight $s=0$. Hence, the polar
solution is simply the spheroidal harmonic $S_s(\theta)={}_{0}\tilde{S}_{\ell
m}(\theta;c)$ associated with the polar eigenvalue $A_{\ell m}(c)$ that reduces
to $A_{\ell m}(c\rightarrow 0)=\ell(\ell+1)$ in the Schwarzschild limit (see,
for instance, Ref.~\cite{Berti:2005gp}). The radial equation turns out to be
the source-free $s=0$ radial Teukolsky equation
\begin{align}
\begin{aligned}
& \frac{d}{dr}\left(\Delta \frac{dR_s}{dr}\right) \\
& \quad +\left(\frac{(r^2+a^2)\omega-am}{\Delta}-\lambda_{\ell m}-\mu^2_Sr^2\right)R_s=0,
\label{eq:radialscalar}
\end{aligned}
\end{align}
where $\lambda_{\ell m}=A_{\ell m}+a^2\omega^2-2am\omega$ depends on the radial eigenvalue $\omega=\omega_R+i\omega_I$, and $\Delta=r^2-2Mr+a^2$. 

The radial eigenvalue $\omega$ can be obtained, together with the radial
solution $R_s(r)$ satisfying ingoing boundary conditions at the horizon, and
asymptotically flat boundary conditions at spatial infinity. At leading order
in $\alpha$, the above radial equation reduces to a type of Laguerre equation,
yielding Hydrogen-like radial states, together with the associated energy
spectrum $\omega$ \cite{Dolan:2007mj,Detweiler:1980uk}. Higher order
corrections at the level of the radial and polar equations are solved for in an
order-by-order fashion perturbatively around $\alpha=0$. Solving the eigenvalue
problem in this way leads to the higher order corrections to the real part of
the superradiantly unstable scalar modes, defined in \eqref{eq:freqnonrel}, \cite{Baumann:2019eav}
\begin{align}
C_S[\alpha]=-\frac{\alpha^4}{8n^4}+\frac{f^S_{n \ell}\alpha^4}{n^3}+\frac{h^S_{\ell} a_* m \alpha^5}{n^3}+\mathcal{O}(\alpha^6),
\label{eq:freqrelcorrectionsscalar}
\end{align}
where 
\begin{align}
\begin{aligned}
f^S_{n\ell}= & \ -\frac{6}{2\ell+1}+\frac{2}{n}, \\
h^S_{\ell}= & \ \frac{16}{2\ell(2\ell+1)(2\ell+2)}.
\end{aligned}
\end{align}
The corresponding instability growth rates, defined in \eqref{eq:ratenonrel}, are \cite{Detweiler:1980uk}
\begin{align}
\begin{aligned}
G_S(a_*,\alpha)= & \ \frac{2^{4\ell+1}(n+\ell)!}{n^{2\ell+4}(n-\ell-1)!}k^S_{n\ell}g^S_{m\ell} \\
k^S_{n\ell}= & \ \left[\frac{\ell!}{(2\ell)!(2\ell+1)!}\right]^2, \\
g^S_{m\ell}= & \ \prod_{o=1}^{\ell}\left[o^2(1-a_*^2)+(a_* m-2r_+\omega_R)^2\right].
\label{eq:raterelcorrectionsscalar}
\end{aligned}
\end{align}
for the most unstable mode in the non-relativistic limit. The principle quantum number $n$ is defined in \eqref{eq:principlequantumnumbers}.

In this work, we compute the eigenvalue $\omega=\omega_R+i\omega_I$ numerically
in the relativistic regime $\mathcal{D}_{\rm int}$ where the analytic methods break
down. The typical approach employed to solve differential eigenvalue problems
of this type goes back to Leaver \cite{Leaver:1985ax}, and was applied to
massive scalar fields in Kerr spacetime in
Refs.~\cite{Dolan:2007mj,Yoshino:2013ofa}. There, the radial solution is
assumed to be written in power series form as 
\begin{align}
R(r)=(r-r_+)^{-i\beta}(r-r_-)^{i\beta+\gamma-1}\sum_{n\geq 0} a_n\left(\frac{r-r_+}{r-r_-}\right)^n,
\label{eq:radialscalarsolution}
\end{align}
with $\beta=2Mr_+(\omega-m\Omega_H)/(r_+-r_-)$ and
$\gamma=M(2\omega^2-\mu^2_S)/k$. Plugging this into the radial equation
\eqref{eq:radialscalar}, one obtains a recurrence relation between the
coefficients $a_n$. This relation is used to obtain a continued fraction
constraint on the frequency $\omega$ for each $\{\ell, m, a, \mu_S\}$. This
constraint is an implicit equation for the eigenvalues $\omega_R$ and
$\omega_I$, which can be solved for numerically using a minimization algorithm
over the complex $\omega$-plane. With the recurrence relation and $\omega$, the
radial solution is constructed using \eqref{eq:radialscalarsolution}. Details
can be found in Refs.~\cite{Dolan:2007mj,Yoshino:2013ofa}. In the
non-relativistic limit, we found that a total number of $N\geq 5000$ terms in the
series expansion is necessary for our desired accuracy, while in the
relativistic regime, a lower number, i.e., $N\leq 1000$, is sufficient. We
construct the spheroidal harmonics ${}_0S_{\ell m}(\theta; c)$ and associated
eigenvalues $A_{\ell m}$ using \texttt{qnm}, a python implementation of a
Leaver-like continued fraction method developed in Ref.~\cite{Stein:2019mop}.

\begin{figure}[t]
\includegraphics[width=0.49\textwidth]{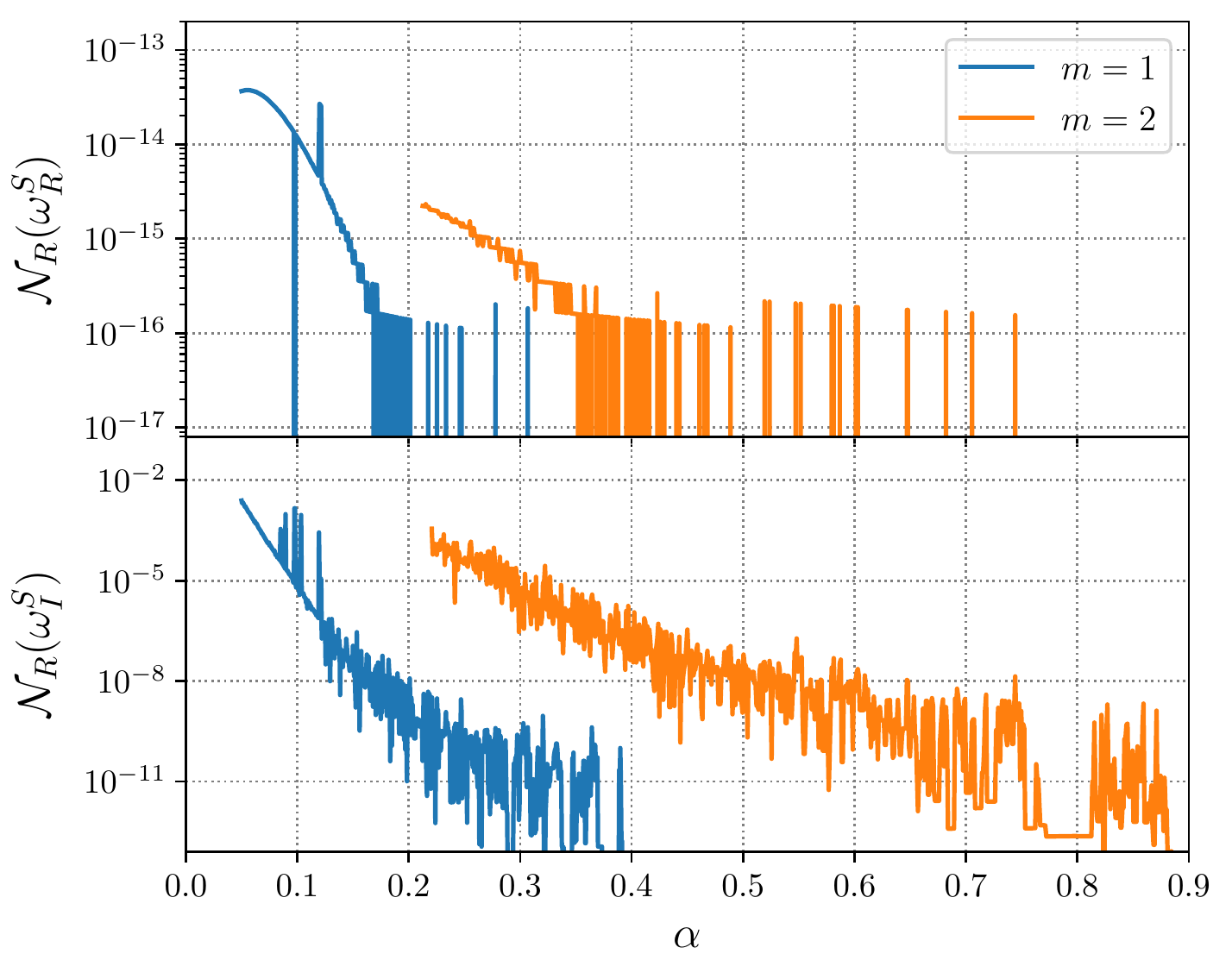}
\caption{The relative numerical error $\mathcal{N}_R$, defined in
\eqref{eq:relerrorscalar}, of the real and imaginary parts of the frequency of
the scalar $m=1$ and $m=2$ superradiant states around a BH of spin
$a_*=0.985$.}
\label{fig:scalarerror}
\end{figure}

In order to estimate the numerical uncertainty of this method, we determine the
frequency $\omega$ in a range of $\alpha$ and fixed BH spin, using the above
approach with successively increasing $N$, up to the $N_{\rm max}=8000$ used
throughout the entire parameter space in \texttt{SuperRad}. The numerical error
is then estimated by
\begin{align}
\mathcal{N}_R(\omega)=\frac{|\omega_{N_{\rm max}}-\omega_{N_{\rm max}/2}|}{\omega_{N_{\rm max}}}.
\label{eq:relerrorscalar}
\end{align}
The results are shown in \figurename{ \ref{fig:scalarerror}}. We ensure that
the minimization algorithm has termination conditions at 
the floating point level. The real part of the
frequency is obtained to one part in $\sim10^{14}$, whereas the
imaginary part is determined less precisely. However, even for $\alpha\gtrsim
0.05$, the latter is more precise or comparable to the theoretical uncertainty
of the analytic estimates in \eqref{eq:freqnonrel} together with
\eqref{eq:freqrelcorrectionsscalar}. This establishes the numerical
uncertainties of the methods used to extend \texttt{SuperRad} into the
relativistic regime. However, it does not show the overall uncertainty of
\texttt{SuperRad} in this regime due to interpolation error, 
which is discussed in Appendix~\ref{app:uncertainties}.

\subsection{Vector field}

The massive vector wave equation \eqref{eq:fieldeq} has been studied more
recently in \cite{Pani:2012bp,Cardoso:2018tly,Dolan:2018dqv,Frolov:2018ezx}.
The non-separability of the vector field equation was a fundamental problem
until a series of works by Lunin \cite{Lunin:2017drx} and Frolov et al.
\cite{Frolov:2018ezx}. There, an ansatz, referred to as FKKS in the following,
was constructed that separates the polar and radial parts of the field equation
\eqref{eq:fieldeq}, and hence, significantly simplifies the problem. We briefly
summarize this ansatz and quote analytic results for the oscillation frequency
and instability growth rates, $\omega=\omega_R+i \omega_I$, obtained with it.
Similarly to the previous subsection, we drop the subscripts of $\omega,\ell,n$,
and $m$, used in the main text, and use these exclusively for vector modes and
frequencies.

The FKKS ansatz exploits a hidden symmetries of Kerr spacetime $g_{\mu\nu}$.
This symmetry is captured by a Killing-Yano 2-form $\textbf{k}$, with tensor
components that satisfy $\nabla_\alpha
k_{\beta\gamma}=2g_{\alpha[\beta}\xi_{\gamma]}$. Using this, the vector field
ansatz takes the form
\begin{align}
A^\mu=B^{\mu\nu}\nabla_\nu Z, & & Z=R_V(r)S_V(\theta)e^{-i(\omega t-m\varphi)},
\end{align}
with polarization tensor $B^{\mu\nu}(g_{\nu\gamma}+i\nu
k_{\nu\gamma})=\delta^\mu_\gamma$ and angular eigenvalue $\nu$. Plugging this
ansatz into \eqref{eq:fieldeq} yields ordinary differential equations for the
radial and polar dependencies, respectively. The angular equation is a deformed
spheroidal harmonic equation for spin-weight $s=-1$ that does not, a priori,
possess known solutions. In the Schwarzschild limit, $a\rightarrow 0$, the
solutions reduce the usual spherical harmonics $S_V(\theta)=Y_{\ell
m}(\theta)$, with a relation between the polar eigenvalue
$\Lambda=\ell(\ell+1)$ and the separation constant $\nu\rightarrow\nu_\ell$
(see Ref.~\cite{Dolan:2018dqv} for details). In this limit, the spatial components of
the vector field are then given by $A^i_{a\rightarrow 0}\propto Y^i_{j,jm}(\theta)$ where 
$Y^i_{j,jm}(\theta)$ are the vector spherical harmonics with $j=\ell-\hat{S}$ (see
Ref.~\cite{Thorne:1980ru}). When the BH spin is non-zero, there is a
mixing of the polar mode number $\ell$, such that, in general
$S_V(\theta)=Y_{|m|,m}(\theta)+b_1 Y_{|m|+1,m}(\theta)+\dots$. The radial
equation for $R_V(r)$ takes the form
\begin{align}
\mathcal{D}_{\nu,\omega, m, a}R_V(r)=0,
\label{eq:procaradialeq}
\end{align}
with a second order differential operator $\mathcal{D}_{\nu,\omega, m, a}$
\cite{Frolov:2018ezx,Dolan:2018dqv}. In the $a\rightarrow 0$ limit, and at
leading order in $\alpha$, this equation reduces to a Schrödinger-type equation
for $R_V(r)$ with the eigenvalue spectrum \eqref{eq:freqnonrel}
\cite{Baryakhtar:2017ngi}. This FKKS ansatz was used in \cite{Baumann:2019eav}
to go beyond the leading orders in both $a$ and $\alpha$. They found the
sub-leading corrections to the spectra \eqref{eq:freqnonrel} to be
\begin{align}
C_V[\alpha]=-\frac{\alpha^4}{8n^4}+\frac{f^V_{n\ell \hat{S}}\alpha^4}{n^3}+\frac{h^V_{\ell \hat{S}} a_* m \alpha^5}{n^3}+\mathcal{O}(\alpha^6),
\label{eq:freqrelcorrectionsvector}
\end{align}
with
\begin{align}
f^V_{n\ell \hat{S}}= & \ -\frac{4(6\ell(\ell-\hat{S}+1)-3\hat{S}+2)}{(2\ell-\hat{S})(2\ell-\hat{S}+1)(2\ell-\hat{S}+2)}+\frac{2}{n},\\
h^V_{\ell \hat{S}}= & \ \frac{16}{(2\ell-\hat{S})(2\ell-\hat{S}+1)(2\ell-\hat{S}+2)}.
\end{align}
The corresponding instability growth rates \eqref{eq:ratenonrel} are \cite{Baryakhtar:2020gao,Baumann:2019eav}
\begin{align}
\begin{aligned}
G_V(a_*,\alpha)= & \ \frac{2^{4\ell-2\hat{S}+1}(n+\ell)!}{n^{2\ell+4}(n-\ell-1)!}k^V_{\ell\hat{S}}d^V_{\ell\hat{S}}g^V_{\ell\hat{S}} \\
k^V_{\ell\hat{S}}= & \ \left[\frac{\ell!}{(2\ell-\hat{S})!(2\ell-\hat{S}+1)!}\right]^2, \\
d^V_{\ell\hat{S}}= & \ \left[1+\frac{2(1+\hat{S})(1-\hat{S})}{2\ell-\hat{S}}\right]^2, \\
g^V_{\ell\hat{S}}= & \ \prod_{o=1}^{\ell-\hat{S}}\left[o^2(1-a_*^2)+(a_* m-2r_+\omega_R)^2\right].
\label{eq:raterelcorrectionsvector}
\end{aligned}
\end{align}
for the most unstable mode in the non-relativistic limit, $\ell=m+\hat{S}$ and $\hat{S}=-1$.

\begin{figure}[t]
\includegraphics[width=0.49\textwidth]{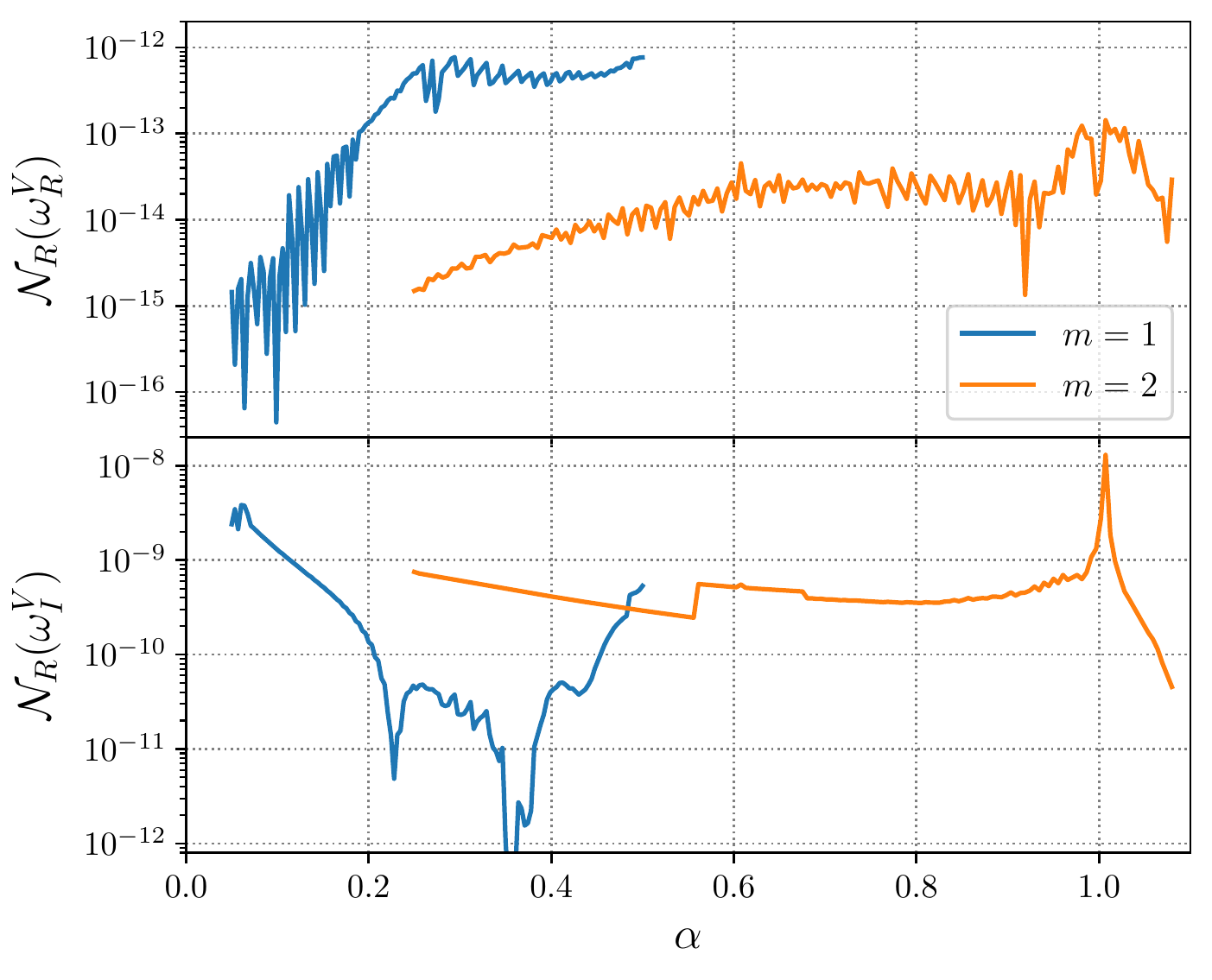}
\caption{The relative numerical error $\mathcal{N}_R$ of the real and imaginary parts of the frequency of the vector $m=1$ and $m=2$ superradiant states around a BH of spin $a_*=0.985$.}
\label{fig:vectorerror}
\end{figure}

In this work, we obtain numerical data in the relativistic regime $\mathcal{D}_{\rm int}$ by
solving \eqref{eq:procaradialeq} and the associated polar equation numerically
following Refs.~\cite{Dolan:2018dqv, Siemonsen:2019ebd}. To that end, the angular
equation is expanded in regular spherical harmonics ${}_0S_{\ell m}(\theta;
c=0)$, while the radial equation is integrated numerically outwards from the
horizon to large distances. Therefore, as long as a sufficient number of terms
is considered in the polar sector, the numerical uncertainties are dominated by
the integration method used in the radial sector. We make use of the
\texttt{BHPToolkit} to construct spherical and spheroidal harmonics
\cite{BHPToolkit}. In order to obtain estimates for the numerical uncertainty,
we vary the step size of the radial numerical integration. In \figurename{
\ref{fig:vectorerror}}, we show upper bounds on the relative numerical
uncertainty of the method described above to obtain the frequencies $\omega$.
As in the scalar case, the numerical uncertainty of the underlying numerical
methods is below the interpolation error of \texttt{SuperRad} discussed in the
next section.

\section{Interpolation and Extrapolation Error of \texttt{SuperRad}} \label{app:uncertainties}

The uncertainties associated to the values of $\tau_I=1/\omega_I$ and
$f_{\rm GW}=\omega_R/\pi$ provided by \texttt{SuperRad} come 
from an interplay of interpolation errors, numerical errors, truncation errors
of analytic expressions, and the theoretical assumptions made. Furthermore, due to
the combination of methods involved, the overall uncertainty of
\texttt{SuperRad} varies across the parameter space. In this appendix, we
provide justifications for accuracy claims made in the main text, as well as
establish upper bounds for uncertainties of the observables contained in
\texttt{SuperRad}.

As we show below, we find the interpolation and extrapolation error to be the
dominant source of error for the waveform model, subdominant to the truncation
error described in the previous section, and shown in \figurename{
\ref{fig:scalarerror}} and \figurename{ \ref{fig:vectorerror}}.  As described
in the main text, and shown in \figurename{ \ref{fig:pspace}}, in the
relativistic regime labelled $\mathcal{D}_{\rm int}$, \texttt{SuperRad} uses linear
interpolation functions
to interpolate based on a grid of $320^2$ data points. We quantify the
interpolation error by directly computing the value of $\omega_R$ and
$\omega_I$ at intermediate value to these data points using the methods outlined in the previous section, and compare them to
the interpolated value. Similarly, we can directly compute the values of
$\omega_R$ and $\omega_I$ in the non-relativistic regime $\mathcal{D}_{\rm fit}$, again using the accurate numerical methods from the previous section, and
compare them to their extrapolated values obtained using the fits.

\begin{figure}[t]
\includegraphics[width=0.49\textwidth]{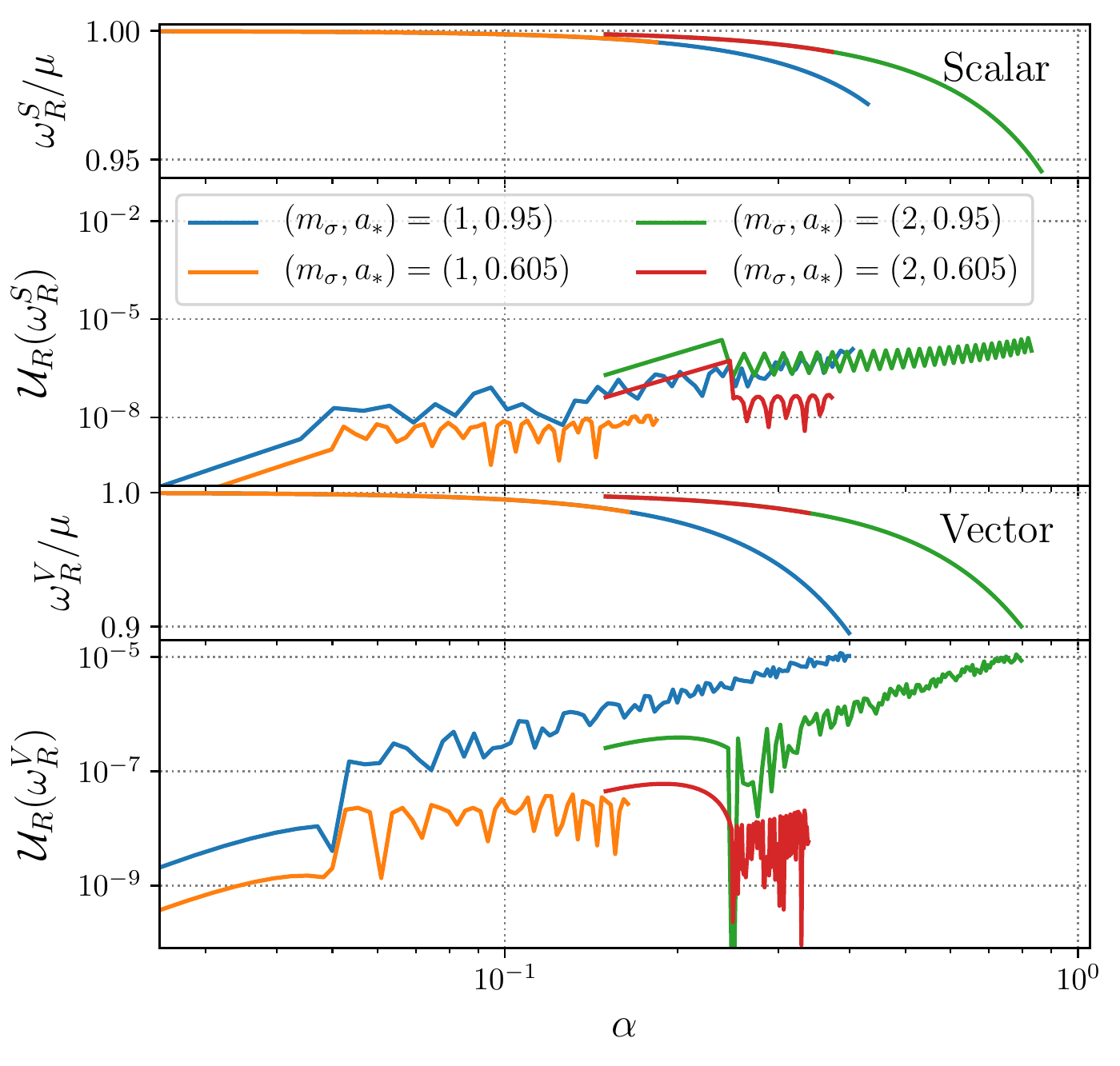}
\caption{We show a set of representative frequencies $\omega_R$ of a $m_\sigma=1$ and $2$ scalar \textit{(top)} and vector \textit{(bottom)} mode (with $\ell_S=m_S$ and $\hat{S}=-1$, respectively), assuming a BH spin of $a_*\in\{0.605,0.95\}$, obtained by \texttt{SuperRad}. We also plot the relative interpolation/extrapolation error $\mathcal{U}_R$ of these predictions (see the main text for discussion).}
\label{fig:frequencieserror}
\end{figure}

\begin{figure}[t]
\includegraphics[width=0.49\textwidth]{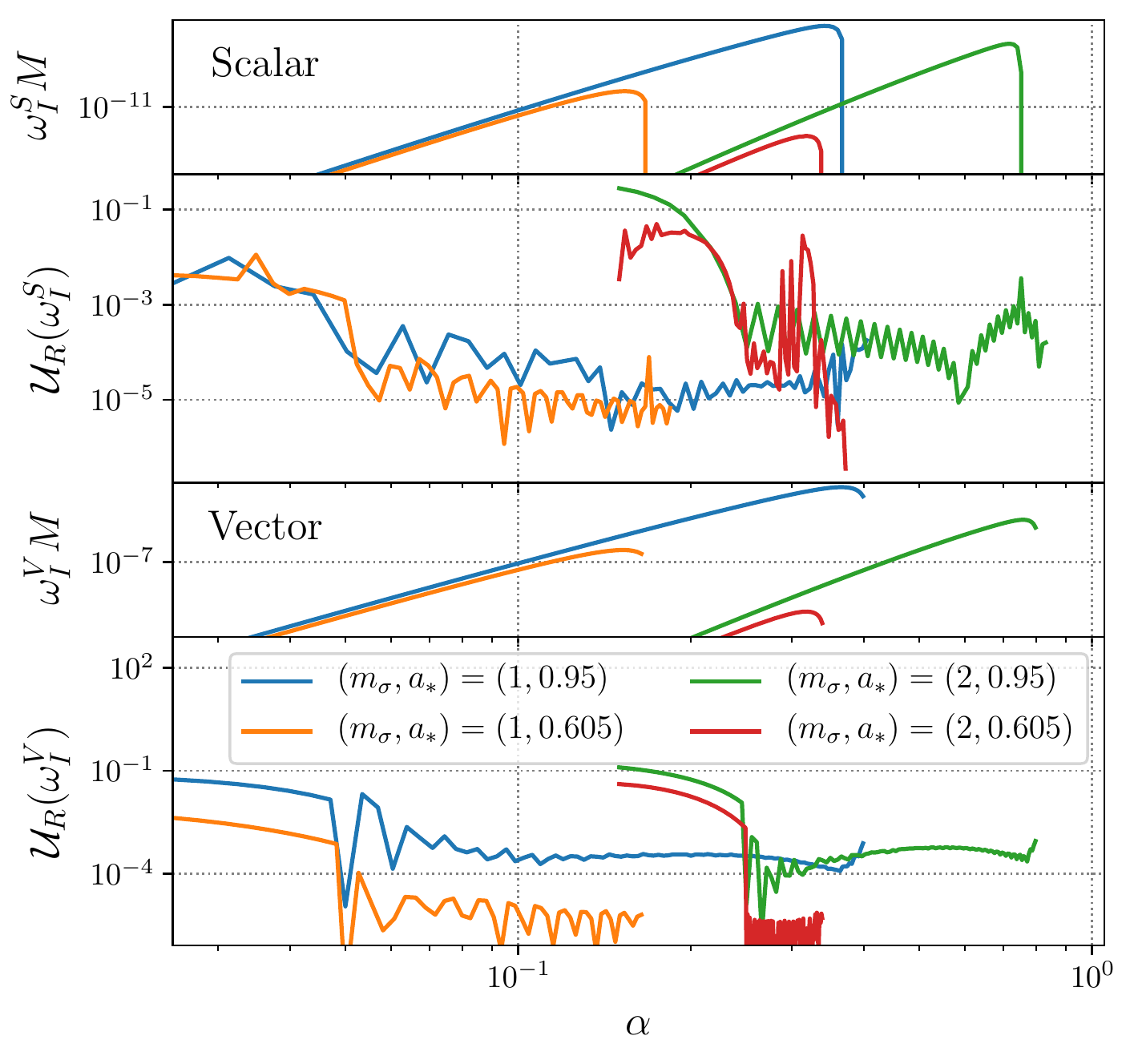}
\caption{We show a set of representative growth rates $\omega_I^\sigma$ of a $m_\sigma=1,2$ scalar \textit{(top)} and vector \textit{(bottom)} mode (with $\ell_S=m_S$ and $\hat{S}=-1$, respectively), assuming a BH spin of $a_*\in\{0.6,0.95\}$, obtained by \texttt{SuperRad}. We also plot the relative interpolation/extrapolation error $\mathcal{U}_R$ of these predictions (see the main text for discussion).}
\label{fig:growthratesserror}
\end{figure}

In \figurename{ \ref{fig:frequencieserror}} and \figurename{
\ref{fig:growthratesserror}}, we show the relative error in interpolating or 
extrapolating $\omega_R$ and $\omega_I$ to a given point in \texttt{SuperRad}, compared to directly computing the value
at that point $\mathcal{U}_R(x)=|x-x_{\rm num.}|/x_{\rm num.}$. 
We show this
for two fiducial spin
values $a_*\in \{0.605,0.95\}$ across the $\alpha$-parameter space from a region
in $\mathcal{D}_{\rm fit}$ to the entire $\mathcal{D}_{\rm int}$ at that spin. The uncertainty
is relatively low in the relativistic regime $\mathcal{D}_{\rm int}$, spanning from
$\alpha^{m_\sigma=1}=0.05$ and $\alpha^{m_\sigma=2}=0.25$ to maximal $\alpha$.
There, the relative uncertainty in the frequency $\omega_R$ does not
exceed $\sim 10^{-5}$, while in the case of the growth rates
$\omega_I$, it is bounded by $\sim 10^{-2}$. In the extrapolated region
$\mathcal{D}_{\rm fit}$, spanning from $\alpha^{m_\sigma=1}=0.05$
and $\alpha^{m_\sigma=2}=0.25$ towards small $\alpha$, the uncertainty
$\mathcal{U}_R(\omega_R)$ of the frequencies is well under control and decreases in the
Newtonian limit. The growth rates, on the other hand, show larger
uncertainties transitioning from $\mathcal{D}_{\rm int}$ towards smaller $\alpha$. The 
fitting procedure, in this case, is more complex, which is reflected in the choices 
of $p$ and $q$ required for \eqref{eq:wifit} (noted below) to be below or comparable 
to the difference between the purely analytic expressions and the numerical 
expressions obtained in parts of $\mathcal{D}_{\rm fit}$. The fit functions and the purely 
analytic estimates for $\omega_I$ have comparable accuracy for $\alpha\lesssim 0.05$ 
and $\alpha\lesssim 0.25$ for $m_\sigma=1$ and $m_\sigma=2$, respectively. The uncertainties are, by construction, decreasing at sufficiently small $\alpha$, while there remains an intermediate regime around $\alpha\approx 0.02$ for $m_\sigma=1$ and $\alpha\approx 0.2$ for $m_\sigma=2$, where the uncertainties first increase. This is a result of the lack of accurate analytical or numerical modeling in this regime. Lastly, since all numerical errors discussed in the previous appendices are
below the uncertainties presented here, the latter can be understood as the
overall uncertainties of the waveform model. Furthermore, comparing the
uncertainties $\mathcal{U}_R$ presented here to the relative differences in
\figurename{ \ref{fig:frequenciesmodelerror}} and  \figurename{
\ref{fig:growthratesmodelerror}}, we see that the latter is always smaller or
comparable to the former. We comment on the uncertainties in the GW emission in
Appendix~\ref{app:gws}. As can be seen in the temporal evolution of the GW frequency emitted by an $m_S=1$ scalar cloud in \figurename{ \ref{fig:strain_freq}}, the quantities (frequencies, timescales, frequency drifts etc.) exhibit a small discontinuity at the interface of $\mathcal{D}_{\rm fit}$ and $\mathcal{D}_{\rm int}$. This is important only, when the system is evolved adiabatically using \eqref{eqn:cloud_evo}, not when the saturation condition $\omega_R=m_\sigma\Omega_H$ is used to set the GW amplitude.

For completeness, we list the different domains $\mathcal{D}_{\rm int}$ used in
\texttt{SuperRad} here. These domains are all bounded by the superradiance
saturation condition $\omega_R=m_\sigma\Omega_H(a_*)$ at sufficiently
large $\alpha$, and by $a_*=0.6$ and $a_*=0.995$. At small $\alpha$ the regions
$\mathcal{D}_{\rm int}$ are bounded by $\alpha_{m_\sigma=1}=0.05$ and
$\alpha_{m_\sigma=2}=0.25$ for both the scalar and the vector clouds. The fit
functions for $\omega_R$ in \eqref{eq:wrfit} contain the following terms: For
$m_V=1$ and $2$ and $m_S=1$ and $2$, we set $q\in\{0,1,\dots,3\}$ and $p\in\{6,7,8\}$. In all four cases,
we added the term $\alpha^5 a_*(\sqrt{1-a_*^2}-1)$. The fit functions for
$\omega_I$ in \eqref{eq:wifit} contain the following terms: For $m_V=1$, we set
$p\in\{1,2,\dots,10\}$, and $q\in\{0,1\}$; for $m_V=2$ we set $p\in\{5,6,\dots,10\}$ and $q\in\{0,1\}$; for
$m_S=1$ we set $p\in\{1,2,3\}$ and $q\in\{0,1,\dots,3\}$; for $m_S=2$ we alter the fit
function slightly with $\hat{c}_{p,q}\rightarrow \delta_{q,1}\hat{c}_p$ and
$\hat{b}_{p,q}\rightarrow \alpha^2 \hat{b}_{p,q}/a_*$ with $p\in\{12,\dots,22\}$ and
$q\in\{0,1,\dots,3\}$. These were fit against the numerical data in $\mathcal{D}_{\rm int}$
with $\alpha<\alpha_{\rm bound}$, where $\alpha_{\rm bound}^{m_\sigma=1}=0.25$
and $\alpha_{\rm bound}^{m_\sigma=2}=0.6$.

\section{Frequency shift} \label{app:frequencyshift}

In this appendix, we briefly discuss the leading-in-$\alpha$ contribution to
the shift in frequency due to the self-gravity of the cloud
$\Delta\omega_\sigma=\alpha^3 M_c F_\sigma/M^2$, where
$F_\sigma$ is defined in \eqref{eq:deltaomegafit}, as described in
Sec.~\ref{sec:freqshift}. For the scalar and vector boson clouds, these are
given by
\begin{align}
\begin{aligned}
	F_S &= \bar{F}[m_S],\\ 
	F_V &= \bar{F}[m_V-1],
\end{aligned}
\label{eq:deltaomegaexpressions}
\end{align}
where
\begin{align}
	\bar{F}[b] = - \frac{2 (b+1)\sqrt{\pi } \Gamma (2 (b+1))-\Gamma \left(2 b+\frac{5}{2}\right)}{2 (b+1)^3\sqrt{\pi } \Gamma (2 (b+1))}.
\label{eq:deltaomegaexpressions2}
\end{align}

\begin{table}[b]
\begin{tabular}{c c | c} 
 $m_V$ & $m_S$ & $M^2\Delta\omega_\sigma/(M_c\alpha^3) $ \\ [0.5ex] 
 \hline\hline
 1 & - & $-5/8$ \\ 
 \hline
 2 & 1 & $-93/512$ \\
 \hline
 3 & 2 & $-793/9216$ \\
 \hline
 4 & 3 & $-26333/524288$ \\
 \hline
 5 & 4 & $-43191/1310720$ \\
 \hline
 6 & 5 & $-1172755/50331648$ \\
\end{tabular}
 \caption{We list the first few leading-in-$\alpha$ contributions to the frequency shift $\Delta\omega_\sigma=\alpha^3 M_c F_\sigma/M^2$ for the oscillation frequency of scalar and vector clouds.}
\label{tab:nonrel_shifts}
\end{table}

In Table~\ref{tab:nonrel_shifts}, we present explicit values for $m_\sigma=1,\dots,5$ for both the scalar and vector cases. The frequency shift monotonically decreases
with increasing $m_\sigma$. The shift depends (to leading order in $\alpha$) only on the $\ell_{\sigma}$ mode number of the considered field,
i.e., the Bohr radius of these non-relativistic solutions, which determines $F_\sigma$, is dependent on the $\ell_{\sigma}$ mode number only.

\section{Gravitational waves} \label{app:gws}

In this appendix, we outline the frequency-domain methods used in the
literature and this work to determine the GWs emitted from a superradiant cloud
after the saturation of the instability, i.e., assuming
$\omega_R=m_\sigma \Omega_H(a_*)$. In the context of the Teukolsky
formalism for linear perturbations on a fixed Kerr spacetime $g_{\mu\nu}$,
finding the GW power and strain reduces to finding the Weyl-Newman-Penrose
scalar $\Psi_4$ at large distances. To that end, the field equations for linear
metric perturbations on $g_{\mu\nu}$ are solved using a separation ansatz
similar to the one used in the previous sections. The polar equation is the
defining equation for spin-weighted spheroidal harmonics, while schematically the sourced
radial Teukolsky equation takes the form \cite{Teukolsky:1973ha}
\begin{align}
\mathcal{D}^{a,M}_{\ell m \omega} R_{\ell m \omega}(r)=\hat{T}_{\ell m \omega}(r),
\label{eq:radialTeukolskyeq}
\end{align}
with sources $\hat{T}_{\ell m \omega}$. In this appendix, $\ell,m$, and $\omega$ refer 
exclusively to modes characterizing the metric perturbations, not the states of the 
superradiant clouds. The second order radial differential
operator $\mathcal{D}^{a,M}_{\ell m \omega}$, for a Kerr spacetime of mass $M$
and spin parameter $a$, is of Sturm-Liouville type and, hence, allows for the generic
construction of a Green's function to solve the inhomogeneous problem 
($\hat{T}_{\ell m \omega}\neq 0$) given the set of homogeneous solutions
$R^H_{\ell m\omega}$ satisfying purely ingoing and purely outgoing boundary
conditions at the horizon, $r=r_+$, and infinity, $r\rightarrow\infty$,
respectively. At large distances $r$, the solution to the radial Teukolsky
equation is
\begin{align}
\begin{aligned}
R_{\ell m\omega}(r\rightarrow\infty)= & \ \frac{r^3e^{i\omega r_*}}{2i\omega B_{\ell m\omega}}\int_{r_+}^\infty dr'\frac{\hat{T}_{\ell m\omega}R^H_{\ell m\omega}}{\Delta^2}\\
= & \ Z^\infty_{\ell m}r^3e^{i\omega r_*}.
\end{aligned}
\label{eq:teukolskyradialsol}
\end{align}
Here, $r_*$ is the Tortoise coordinate of $r$, and we defined a set of variables
$Z^\infty_{\ell m}$ containing information about the source (see, e.g.,
Ref.~\cite{Siemonsen:2019ebd}, for details). With this in hand, the GWs at infinity can be
calculated as
\begin{align}
\Psi_4=\frac{1}{r}\sum_{\ell,m}\frac{Z^\infty_{\ell m}}{\sqrt{2\pi}}e^{i\omega(r_*-t)} {}_{-2}S_{\ell m}(\theta;c=a\omega)e^{im\varphi} .
\label{eq:Teukolskypsi4}
\end{align}
Notice, the summation in \eqref{eq:Teukolskypsi4} is over \textit{spheroidal}
$\ell$, rather than \textit{spherical} $\ell$ as done in
Sec.~\ref{sec:conventions}. Here, we are using the normalization $\int
d\cos\theta |{}_{-2}S_{\ell m}(\theta;c)|^2=1$. To recover the spherical
harmonic GW modes $h^{\ell'm'}$, we rewrite the above, using
\begin{align}
h=-\frac{2\Psi_4}{(2\omega_R)^2},
\label{eq:relhtopsi}
\end{align}
leading to
\begin{align}
r e^{i\omega t}h^{\ell' m'}=\sum_{\ell,m}\frac{-2Z^\infty_{\ell m}}{\sqrt{2\pi}(2\omega_R)^2}C^{\ell' m'}_{\ell m}=\frac{-2\tilde{Z}^\infty_{\ell'm'}}{\sqrt{2\pi}(2\omega_R)^2},
\label{eq:gwmodesfromteukolsky}
\end{align}
with
\begin{align}
C^{\ell' m'}_{\ell m}=\int_{S^2}d\Omega {}_{-2}\bar{Y}_{\ell' m'}(\Omega){}_{-2}S_{\ell m}(\theta;c=a\omega)e^{im\varphi}.
\end{align}
Note that when $c=0$, $C^{\ell' m'}_{\ell m}=2 \pi \delta^{\ell'}_{\ell} \delta^{m'}_{m}$. The total emitted gravitational energy flux is
\begin{align}
P_{\rm GW}=\sum_{\ell',m'}\frac{|\tilde{Z}^\infty_{\ell'm'}|^2}{8\pi^2(2\omega_R)^2}.
\end{align}
Therefore, determining the GWs emitted depends on finding homogeneous solutions
to the radial Teukolsky equation, as well as integrating these over the cloud
sources $\hat{T}_{\ell m\omega}$. The three distinct approximations mentioned
in the main text---flat, Schwarzschild, and Teukolsky---all emerge from
\eqref{eq:teukolskyradialsol} by dropping certain terms. In the flat
approximation, the spin is neglected, $a=0$, and the source equations are
expanded to leading order in $\alpha$, implying $M\rightarrow 0$. In this
limit, both the homogeneous solutions and source functions can be constructed
and integrated over analytically. In the Schwarzschild approximation, one also expands in
$\alpha$ to leading order and assumes $a=0$. However, one includes the
gravitational potential terms present in the Schwarzschild Green's function,
i.e., $M\neq 0$. For $\ell'=m'=2$, the flat ansatz generally
underestimates the emitted GW flux, while the Schwarzschild approximation
overestimates the power. Solving the equation \eqref{eq:radialTeukolskyeq}
numerically making no assumptions about $\alpha$ or $a$ (referred to as the
Teukolsky approximation in the main text) provides the most accurate
predictions for $P_{\rm GW}$ and $h$, and is expected to give values intermediate to the
flat and the Schwarzschild approximations. More details can be found in,
for instance, Refs.~\cite{brito,Baryakhtar:2017ngi}.

For $\ell'=m'>2$, i.e., $m_\sigma>1$, the GW energy flux has
been computed analytically only in the flat approximation. In the scalar
case, the total GW power emitted from a cloud with $(n_S,m_S)$ and $\ell_S=m_S$, is given
by \cite{Yoshino:2013ofa}
\begin{align}
P_{\rm GW}=C_{n_S m_S}\alpha^{Q_S}\frac{M_c^2}{M^2},
\end{align}
where $Q_S=4m_S+10$ and 
\begin{align}
\begin{aligned}
C_{n_S m_S}=\frac{16^{m_S+1}m_S(2m_S-1)}{n_S^{4m_S+8}(m_S+1)\Gamma(m_S+1)^4}\\
\times \frac{\Gamma(2m_S-1)^2\Gamma(m_S+n_S+1)^2}{\Gamma(4m_S+3)\Gamma(n_S-m_S)^2}.
\end{aligned}
\end{align}
In the vector case, for $m'>1$, the GW power emitted from a $\hat{S}=-1$ superradiant state in the $\alpha\ll 1$ limit is \cite{Baryakhtar:2017ngi}
\begin{align}
P_{\rm GW}=K_{m_V} \alpha^{Q_V} \frac{M_c^2}{M^2},
\end{align}
where $Q_V=4m_V+6$, $K_2=1/126, K_3=6\times 10^{-6}, K_4=2\times 10^{-9}$, and $K_5=4\times 10^{-13}$.

\begin{figure}[t]
\includegraphics[width=0.49\textwidth]{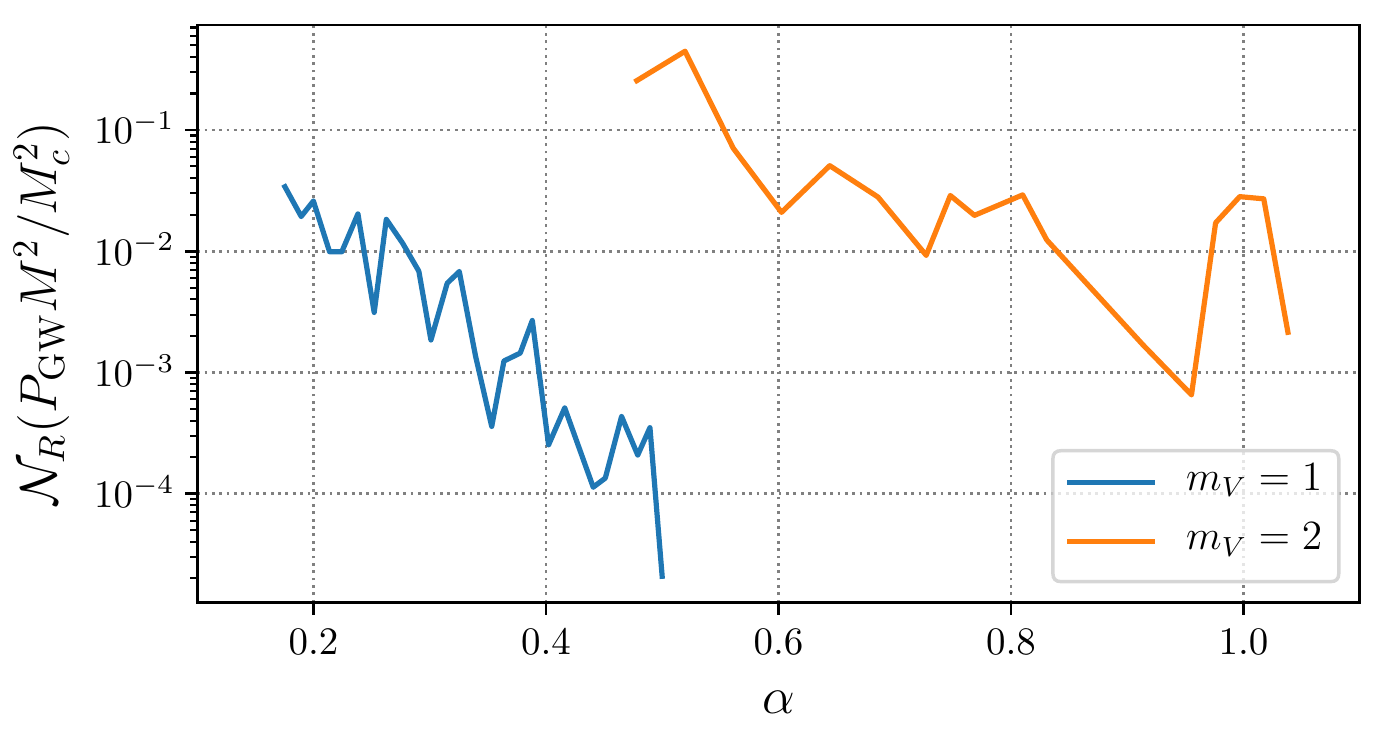}
\caption{The relative numerical error $\mathcal{N}_R$ of the total emitted
GW energy flux $P_{\rm GW}$ from a vector cloud in the $m_V=1$
and $m_V=2$ superradiant states around a BH of spin $a=0.985M$ in the relevant
part of the parameter space.}
\label{fig:powererror}
\end{figure}

The specifics of the methods we use to numerically solve
\eqref{eq:radialTeukolskyeq} are discussed in detail in
Ref.~\cite{Siemonsen:2019ebd} (we make use of the \texttt{BHPToolkit}
\cite{BHPToolkit}). It involves constructing the sources $\hat{T}_{\ell
m\omega}$ from the numerical superradiant solutions to \eqref{eq:fieldeq}, and
integrating \eqref{eq:teukolskyradialsol} numerically. In \figurename{
\ref{fig:powererror}}, we present upper bounds on the numerical error of these
methods across the entire parameter space, assuming a $m_V=1$ and $2$ vector
cloud (analogous upper bounds are expected for scalar clouds). The bounds are
obtained from varying the resolution of the underlying superradiant vector
field solution together with the radial step size of the numerical integration
of \eqref{eq:teukolskyradialsol}. The relative difference between estimates of
$\tilde{P}_{\rm GW}$ with two different resolutions decreases with increasing
resolution. The upper bounds shown in \figurename{ \ref{fig:powererror}} are
the relative difference between the default resolution used throughout, and
half that resolution. As
for the GW power calculation, the GW strain $h$ is calculated from the
solutions to \eqref{eq:teukolskyradialsol}, through
\eqref{eq:gwmodesfromteukolsky}. Since $h\sim \sqrt{P_{\rm GW}}$, the values for
$\mathcal{N}_R (\tilde{P}_{\rm GW})$ can be interpreted as an estimate for
the error $\mathcal{N}_R(h_{\times,+}M/M_c)$ for the amplitudes of the polarization
waveform. 

Lastly, the numerical data regimes $\tilde{D}_{\rm int}$ (defined in Sec.~\ref{sec:gwpowerstrain}) is bounded at large $\alpha$ by the maximal $\alpha$
satisfying the superradiance saturation condition $\omega_R=m_\sigma\Omega_H$
at the corresponding spin $a_*$. From below, it is bounded by $\alpha_{\rm
low}^{m_S=1}=0.2$ and $\alpha_{\rm low}^{m_S=2}=0.34$, for scalars and the two
lowest azimuthal numbers, and $\alpha_{\rm low}^{m_V=1}=0.17$ and $\alpha_{\rm
low}^{m_V=2}=0.45$, for vectors with the two lowest azimuthal numbers. 

\bibliography{bib.bib,bib2.bib}

\end{document}